\input harvmac
\input OpenClosedV1.defs
\noblackbox

\input epsf
\input graphicx
  

\input color 
     
  
 
\def\journal#1&#2(#3){\unskip, \sl #1\ \bf #2 \rm(19#3) }
\def\andjournal#1&#2(#3){\sl #1~\bf #2 \rm (19#3) }

\def\frac#1#2{{#1\over#2}}

\def\d{\partial}

\def\inbar{\,\vrule height1.5ex width.4pt depth0pt}
\def\IC{\relax\hbox{$\inbar\kern-.3em{\rm C}$}}
\def\IR{\relax{\rm I\kern-.18em R}}
\def\IP{\relax{\rm I\kern-.18em P}}
\def\IZ{\relax{\rm I\kern-.18em Z}}
\def\IE{\relax{\rm I\kern-.18em E}}

\def\dd{\relax{$\delta$\kern-.43em $\delta$}}
\def\dde{\relax{\delta\kern-0.43em \delta}}
%
%

%
\catcode`\@=11
\def\slash#1{\mathord{\mathpalette\c@ncel{#1}}}
\overfullrule=0pt
\def\AA{{\cal A}}
\def\BB{{\cal B}}
\def\CC{{\cal C}}

\def\FF{{\cal F}}

\def\II{{\cal I}}
\def\JJ{{\cal J}}

\def\LL{{\cal L}}

\def\OO{{\cal O}}

\def\RR{{\cal R}}
\def\SS{{\cal S}}
\def\TT{{\cal T}}

\def\VV{{\cal V}}

\def\ZZ{{\cal Z}}

\def\underrel#1\over#2{\mathrel{\mathop{\kern\z@#1}\limits_{#2}}}

\catcode`\@=12


%

\def\det{{\rm det}}

\def \sinh{{\rm sinh}}

\def\det{{\rm det}}


\def\D{{\partial}}


\def\unlockat{\catcode`\@=11}
\def\lockat{\catcode`\@=12}

\unlockat


\def\newsec#1{\global\advance\secno by1\message{(\the\secno. #1)}
\global\subsecno=0\global\subsubsecno=0\eqnres@t\noindent
{\bf\the\secno. #1}
\writetoca{{\secsym} {#1}}\par\nobreak\medskip\nobreak}
\global\newcount\subsecno \global\subsecno=0
\def\subsec#1{\global\advance\subsecno
by1\message{(\secsym\the\subsecno. #1)}
\ifnum\lastpenalty>9000\else\bigbreak\fi\global\subsubsecno=0
\noindent{\it\secsym\the\subsecno. #1}
\writetoca{\string\quad {\secsym\the\subsecno.} {#1}}
\par\nobreak\medskip\nobreak}
\global\newcount\subsubsecno \global\subsubsecno=0
\def\subsubsec#1{\global\advance\subsubsecno by1
\message{(\secsym\the\subsecno.\the\subsubsecno. #1)}
\ifnum\lastpenalty>9000\else\bigbreak\fi
\noindent\quad{\secsym\the\subsecno.\the\subsubsecno.}{#1}
\writetoca{\string\qquad{\secsym\the\subsecno.\the\subsubsecno.}{#1}}
\par\nobreak\medskip\nobreak}

\def\subsubseclab#1{\DefWarn#1\xdef
#1{\noexpand\hyperref{}{subsubsection}%
{\secsym\the\subsecno.\the\subsubsecno}%
{\secsym\the\subsecno.\the\subsubsecno}}%
\writedef{#1\leftbracket#1}\wrlabeL{#1=#1}}
\def\seclab#1{\xdef #1{\the\secno}\writedef{#1\leftbracket#1}\wrlabeL{#1=#1}}
\def\subseclab#1{\xdef #1{\secsym\the\subsecno}%
\writedef{#1\leftbracket#1}\wrlabeL{#1=#1}}
\lockat


\newcount\figno
\figno=1
\def\fig#1#2#3{
\par\begingroup\parindent=0pt\leftskip=1cm\rightskip=1cm\parindent=0pt
\baselineskip=11pt
\global\advance\figno by 1
\midinsert
\epsfxsize=#3
\centerline{\epsfbox{#2}}
{\bf Fig.\ \the\figno: } #1\par
\endinsert\endgroup\par
}
\def\figlabel#1{\xdef#1{\the\figno}}
\def\encadremath#1{\vbox{\hrule\hbox{\vrule\kern8pt\vbox{\kern8pt
\hbox{$\displaystyle #1$}\kern8pt}
\kern8pt\vrule}\hrule}}
%
%


\font\cmss=cmss10
\font\cmsss=cmss10 at 7pt
\def\rlx{\relax\leavevmode}
\def\inbar{\vrule height1.5ex width.4pt depth0pt}
\def\IC{\relax\,\hbox{$\inbar\kern-.3em{\rm C}$}}
\def\IN{\relax{\rm I\kern-.18em N}}
\def\IP{\relax{\rm I\kern-.18em P}}
\def\ZZ{\rlx\leavevmode\ifmmode\mathchoice{\hbox{\cmss Z\kern-.4em Z}}
 {\hbox{\cmss Z\kern-.4em Z}}{\lower.9pt\hbox{\cmsss Z\kern-.36em Z}}
 {\lower1.2pt\hbox{\cmsss Z\kern-.36em Z}}\else{\cmss Z\kern-.4em
 Z}\fi}
\def\IZ{\relax\ifmmode\mathchoice
{\hbox{\cmss Z\kern-.4em Z}}{\hbox{\cmss Z\kern-.4em Z}}
{\lower.9pt\hbox{\cmsss Z\kern-.4em Z}}
{\lower1.2pt\hbox{\cmsss Z\kern-.4em Z}}\else{\cmss Z\kern-.4em
Z}\fi}
\def\IZ{\relax\ifmmode\mathchoice
{\hbox{\cmss Z\kern-.4em Z}}{\hbox{\cmss Z\kern-.4em Z}}
{\lower.9pt\hbox{\cmsss Z\kern-.4em Z}}
{\lower1.2pt\hbox{\cmsss Z\kern-.4em Z}}\else{\cmss Z\kern-.4em
Z}\fi}

\def\narrowplus{\kern -.04truein + \kern -.03truein}
\def\narrowminus{- \kern -.04truein}
\def\narrowminussub{\kern -.02truein - \kern -.01truein}
\font\ninerm=cmr9

\def\IZ{\relax\ifmmode\mathchoice
{\hbox{\cmss Z\kern-.4em Z}}{\hbox{\cmss Z\kern-.4em Z}}
{\lower.9pt\hbox{\cmsss Z\kern-.4em Z}}
{\lower1.2pt\hbox{\cmsss Z\kern-.4em Z}}\else{\cmss Z\kern-.4em
Z}\fi}
\def\IB{\relax{\rm I\kern-.18em B}}
\def\IC{{\relax\hbox{$\inbar\kern-.3em{\rm C}$}}}
\def\ID{\relax{\rm I\kern-.18em D}}
\def\IE{\relax{\rm I\kern-.18em E}}
\def\IF{\relax{\rm I\kern-.18em F}}
\def\IG{\relax\hbox{$\inbar\kern-.3em{\rm G}$}}
\def\IGa{\relax\hbox{${\rm I}\kern-.18em\Gamma$}}
\def\IH{\relax{\rm I\kern-.18em H}}
\def\II{\relax{\rm I\kern-.18em I}}
\def\IK{\relax{\rm I\kern-.18em K}}
\def\IP{\relax{\rm I\kern-.18em P}}

\font\cmss=cmss10 \font\cmsss=cmss10 at 7pt
\def\IR{\relax{\rm I\kern-.18em R}}


%

%
%
\def\eqnn#1{\xdef #1{(\secsym\the\meqno)}\writedef{#1\leftbracket#1}%
\global\advance\meqno by1\wrlabeL#1}
\def\eqna#1{\xdef #1##1{\hbox{$(\secsym\the\meqno##1)$}}
\writedef{#1\numbersign1\leftbracket#1{\numbersign1}}%
\global\advance\meqno by1\wrlabeL{#1$\{\}$}}
\def\eqn#1#2{\xdef #1{(\secsym\the\meqno)}\writedef{#1\leftbracket#1}%
\global\advance\meqno by1$$#2\eqno#1\eqlabeL#1$$}


\def\boxit#1{\vbox{\hrule\hbox{\vrule\kern8pt
\vbox{\hbox{\kern8pt}\hbox{\vbox{#1}}\hbox{\kern8pt}}
\kern8pt\vrule}\hrule}}
\def\mathboxit#1{\vbox{\hrule\hbox{\vrule\kern5pt\vbox{\kern5pt
\hbox{$\displaystyle #1$}\kern5pt}\kern5pt\vrule}\hrule}}


\lref\EmparanCS{
  R.~Emparan, T.~Harmark, V.~Niarchos and N.~A.~Obers,
  ``World-Volume Effective Theory for Higher-Dimensional Black Holes,''
Phys.\ Rev.\ Lett.\  {\bf 102}, 191301 (2009).
[arXiv:0902.0427 [hep-th]].
}

\lref\EmparanHG{
  R.~Emparan, T.~Harmark, V.~Niarchos and N.~A.~Obers,
  ``Blackfolds in Supergravity and String Theory,''
JHEP {\bf 1108}, 154 (2011).
[arXiv:1106.4428 [hep-th]].
}

\lref\EmparanAT{
  R.~Emparan, T.~Harmark, V.~Niarchos and N.~A.~Obers,
  ``Essentials of Blackfold Dynamics,''
  JHEP {\bf 1003} (2010) 063
  [arXiv:0910.1601 [hep-th]].
}

\lref\GrignaniXM{
  G.~Grignani, T.~Harmark, A.~Marini, N.~A.~Obers and M.~Orselli,
  ``Heating up the BIon,''
JHEP {\bf 1106}, 058 (2011).
[arXiv:1012.1494 [hep-th]].
}

\lref\GrignaniXMM{
  G.~Grignani, T.~Harmark, A.~Marini, N.~A.~Obers and M.~Orselli,
  ``Thermodynamics of the hot BIon,''
  Nucl.\ Phys.\ B {\bf 851} (2011) 462
  [arXiv:1101.1297 [hep-th]].
}

\lref\CampsHW{
  J.~Camps and R.~Emparan,
  ``Derivation of the blackfold effective theory,''
JHEP {\bf 1203}, 038 (2012).
[arXiv:1201.3506 [hep-th]].
}

\lref\Simon{
  J.~Simon,
  ``Brane Effective Actions, Kappa-Symmetry and Applications,''
  Living Rev.\ Rel.\  {\bf 5} (2012) 3
  [arXiv:1110.2422 [hep-th]].
}

\lref\NiarchosCY{
  V.~Niarchos and K.~Siampos,
  ``Entropy of the self-dual string soliton,''
JHEP {\bf 1207}, 134 (2012).
[arXiv:1206.2935 [hep-th]].
}

\lref\NiarchosPN{
  V.~Niarchos and K.~Siampos,
  ``M2-M5 blackfold funnels,''
JHEP {\bf 1206}, 175 (2012).
[arXiv:1205.1535 [hep-th]].
}

\lref\CampsBR{
  J.~Camps, R.~Emparan and N.~Haddad,
  ``Black Brane Viscosity and the Gregory-Laflamme Instability,''
JHEP {\bf 1005}, 042 (2010).
[arXiv:1003.3636 [hep-th]].
}

\lref\FluxBlackfolds{
 J.~ Armas, J.~ Gath, V.~Niarchos, N.~A.~Obers and A.~V.~Pedersen,
  to appear.
}

\lref\NiarchosIA{
  V.~Niarchos and K.~Siampos,
  ``The black M2-M5 ring intersection spins,''
PoS Corfu {\bf 2012}, 088 (2013).
[arXiv:1302.0854 [hep-th]].
}

\lref\BhattacharyyaJC{
  S.~Bhattacharyya, V.~E.~Hubeny, S.~Minwalla and M.~Rangamani,
  ``Nonlinear Fluid Dynamics from Gravity,''
JHEP {\bf 0802}, 045 (2008).
[arXiv:0712.2456 [hep-th]].
}

\lref\GathQYA{
  J.~Gath and A.~V.~Pedersen,
  ``Viscous Asymptotically Flat Reissner-Nordstr\"om Black Branes,''
[arXiv:1302.5480 [hep-th]].
}

\lref\EmparanUX{
  R.~Emparan, D.~Mateos and P.~K.~Townsend,
  ``Supergravity supertubes,''
JHEP {\bf 0107}, 011 (2001).
[hep-th/0106012].
}

\lref\LuninMJ{
  O.~Lunin,
  ``Strings ending on branes from supergravity,''
JHEP {\bf 0709}, 093 (2007).
[arXiv:0706.3396 [hep-th]].
}

\lref\LuninTF{
  O.~Lunin,
  ``Brane webs and 1/4-BPS geometries,''
JHEP {\bf 0809}, 028 (2008).
[arXiv:0802.0735 [hep-th]].
}

\lref\TseytlinDJ{
  A.~A.~Tseytlin,
  ``Born-Infeld action, supersymmetry and string theory,''
In *Shifman, M.A. (ed.): The many faces of the superworld* 417-452.
[hep-th/9908105].
}

\lref\MyersBW{
  R.~C.~Myers,
  ``NonAbelian phenomena on D branes,''
Class.\ Quant.\ Grav.\  {\bf 20}, S347 (2003).
[hep-th/0303072].
}

\lref\EmparanILA{
  R.~Emparan, V.~E.~Hubeny and M.~Rangamani,
  ``Effective hydrodynamics of black D3-branes,''
JHEP {\bf 1306}, 035 (2013).
[arXiv:1303.3563 [hep-th]].
}

\lref\GrignaniEWA{
  G.~Grignani, T.~Harmark, A.~Marini and M.~Orselli,
  ``Thermal DBI action for the D3-brane at weak and strong coupling,''
[arXiv:1311.3834 [hep-th]].
}

\lref\GrignaniIW{
  G.~Grignani, T.~Harmark, A.~Marini, N.~A.~Obers and M.~Orselli,
  ``Thermal string probes in AdS and finite temperature Wilson loops,''
JHEP {\bf 1206}, 144 (2012).
[arXiv:1201.4862 [hep-th]].
}

\lref\ArmasBK{
  J.~Armas, T.~Harmark, N.~A.~Obers, M.~Orselli and A.~V.~Pedersen,
  ``Thermal Giant Gravitons,''
JHEP {\bf 1211}, 123 (2012).
[arXiv:1207.2789 [hep-th]].
}

\lref\EmparanVD{
  R.~Emparan, T.~Harmark, V.~Niarchos and N.~A.~Obers,
  ``New Horizons for Black Holes and Branes,''
JHEP {\bf 1004}, 046 (2010).
[arXiv:0912.2352 [hep-th]].
}

\lref\CampsHB{
  J.~Camps, R.~Emparan, P.~Figueras, S.~Giusto and A.~Saxena,
  ``Black Rings in Taub-NUT and D0-D6 interactions,''
JHEP {\bf 0902}, 021 (2009).
[arXiv:0811.2088 [hep-th]].
}

\lref\ArmasHZ{
  J.~Armas and N.~A.~Obers,
  ``Blackfolds in (Anti)-de Sitter Backgrounds,''
Phys.\ Rev.\ D {\bf 83}, 084039 (2011).
[arXiv:1012.5081 [hep-th]].
}

\lref\ArmasOTA{
  J.~Armas, N.~A.~Obers and A.~V.~Pedersen,
  ``Null-Wave Giant Gravitons from Thermal Spinning Brane Probes,''
JHEP {\bf 1310}, 109 (2013).
[arXiv:1306.2633 [hep-th]].
}

\lref\ArmasAKA{
  J.~Armas, J.~Gath and N.~A.~Obers,
  ``Electroelasticity of Charged Black Branes,''
JHEP {\bf 1310}, 035 (2013).
[arXiv:1307.0504 [hep-th]].
}

\lref\GauntlettDI{
  J.~P.~Gauntlett,
  ``Branes, calibrations and supergravity,''
[hep-th/0305074].
}

\lref\CallanKY{
  C.~G.~Callan, Jr., J.~A.~Harvey and A.~Strominger,
  ``Worldbrane actions for string solitons,''
Nucl.\ Phys.\ B {\bf 367}, 60 (1991).
}

\lref\GibbonsSV{
  G.~W.~Gibbons and P.~K.~Townsend,
  ``Vacuum interpolation in supergravity via super p-branes,''
Phys.\ Rev.\ Lett.\  {\bf 71}, 3754 (1993).
[hep-th/9307049].
}

\lref\CallanKZ{
  C.~G.~Callan and J.~M.~Maldacena,
  ``Brane death and dynamics from the Born-Infeld action,''
Nucl.\ Phys.\ B {\bf 513}, 198 (1998).
[hep-th/9708147].
}

\lref\TseytlinWW{
  A.~A.~Tseytlin,
  ``Renormalization of Mobius Infinities and Partition Function Representation for String Theory Effective Action,''
Phys.\ Lett.\ B {\bf 202}, 81 (1988).
}

\lref\AndreevCB{
  O.~D.~Andreev and A.~A.~Tseytlin,
  ``Partition Function Representation for the Open Superstring Effective Action: Cancellation of 
  Mobius Infinities and Derivative Corrections to Born-Infeld Lagrangian,''
Nucl.\ Phys.\ B {\bf 311}, 205 (1988).
}

\lref\AbouelsaoodGD{
  A.~Abouelsaood, C.~G.~Callan, Jr., C.~R.~Nappi and S.~A.~Yost,
  ``Open Strings in Background Gauge Fields,''
Nucl.\ Phys.\ B {\bf 280}, 599 (1987).
}

\lref\AndreevRE{
  O.~D.~Andreev and A.~A.~Tseytlin,
  ``Two Loop Beta Function in the Open String $\sigma$ Model and 
  Equivalence With String Effective Equations of Motion,''
Mod.\ Phys.\ Lett.\ A {\bf 3}, 1349 (1988).
}

\lref\AnderssonNR{
  N.~Andersson and G.~L.~Comer,
  ``Relativistic fluid dynamics: Physics for many different scales,''
Living Rev.\ Rel.\  {\bf 10}, 1 (2007).
[gr-qc/0605010].
}

\lref\hydrobook{
  L.~Rezzolla and O.~Zanotti,
  ``Relativistic Hydrodynamics,''
  Oxford University Press (2013),
  DOI:10.1093/acprof:oso/9780198528906.001.0001
}

\lref\SenIV{
  A.~Sen,
  ``Open closed duality: Lessons from matrix model,''
Mod.\ Phys.\ Lett.\ A {\bf 19}, 841 (2004).
[hep-th/0308068].
}

\lref\SenNF{
  A.~Sen,
  ``Tachyon dynamics in open string theory,''
Int.\ J.\ Mod.\ Phys.\ A {\bf 20}, 5513 (2005).
[hep-th/0410103].
}

\lref\BondiPX{
  H.~Bondi, M.~G.~J.~van der Burg and A.~W.~K.~Metzner,
  ``Gravitational waves in general relativity. 7. Waves from axisymmetric isolated systems,''
Proc.\ Roy.\ Soc.\ Lond.\ A {\bf 269}, 21 (1962).
}

\lref\SachsZZA{
  R.~Sachs,
  ``Asymptotic symmetries in gravitational theory,''
Phys.\ Rev.\  {\bf 128}, 2851 (1962).
}

\lref\ArcioniXX{
  G.~Arcioni and C.~Dappiaggi,
  ``Exploring the holographic principle in asymptotically flat space-times via the BMS group,''
Nucl.\ Phys.\ B {\bf 674}, 553 (2003).
[hep-th/0306142].
}

\lref\ArcioniTD{
  G.~Arcioni and C.~Dappiaggi,
  ``Holography in asymptotically flat space-times and the BMS group,''
Class.\ Quant.\ Grav.\  {\bf 21}, 5655 (2004).
[hep-th/0312186].
}

\lref\DappiaggiCI{
  C.~Dappiaggi, V.~Moretti and N.~Pinamonti,
  ``Rigorous steps towards holography in asymptotically flat spacetimes,''
Rev.\ Math.\ Phys.\  {\bf 18}, 349 (2006).
[gr-qc/0506069].
}

\lref\BarnichEB{
  G.~Barnich and C.~Troessaert,
  ``Aspects of the BMS/CFT correspondence,''
JHEP {\bf 1005}, 062 (2010).
[arXiv:1001.1541 [hep-th]].
}

\lref\deBoerVF{
  J.~de Boer and S.~N.~Solodukhin,
  ``A Holographic reduction of Minkowski space-time,''
Nucl.\ Phys.\ B {\bf 665}, 545 (2003).
[hep-th/0303006].
}

\lref\SolodukhinGS{
  S.~N.~Solodukhin,
  ``Reconstructing Minkowski space-time,''
[hep-th/0405252].
}

\lref\NiarchosMAA{
  V.~Niarchos,
  ``Supersymmetric Perturbations of the M5 brane,''
JHEP {\bf 1405}, 023 (2014).
[arXiv:1402.4132 [hep-th]].
}

\lref\MaldacenaRE{
  J.~M.~Maldacena,
  ``The Large N limit of superconformal field theories and supergravity,''
Int.\ J.\ Theor.\ Phys.\  {\bf 38}, 1113 (1999), [Adv.\ Theor.\ Math.\ Phys.\  {\bf 2}, 231 (1998)].
[hep-th/9711200].
}

\lref\HaehlPJA{
  F.~M.~Haehl, R.~Loganayagam and M.~Rangamani,
  ``Adiabatic hydrodynamics: The eightfold way to dissipation,''
JHEP {\bf 1505}, 060 (2015).
[arXiv:1502.00636 [hep-th]].
}

\lref\deBoerIJA{
  J.~de Boer, M.~P.~Heller and N.~Pinzani-Fokeeva,
  ``Effective actions for relativistic fluids from holography,''
[arXiv:1504.07616 [hep-th]].
}

\lref\DubovskySJ{
  S.~Dubovsky, L.~Hui, A.~Nicolis and D.~T.~Son,
  ``Effective field theory for hydrodynamics: thermodynamics, and the derivative expansion,''
Phys.\ Rev.\ D {\bf 85}, 085029 (2012).
[arXiv:1107.0731 [hep-th]].
}

\lref\GiddingsWP{
  S.~B.~Giddings, E.~J.~Martinec and E.~Witten,
  ``Modular Invariance in String Field Theory,''
Phys.\ Lett.\ B {\bf 176}, 362 (1986).
}

\lref\ZwiebachAZ{
  B.~Zwiebach,
  ``A Proof that Witten's open string theory gives a single cover of moduli space,''
Commun.\ Math.\ Phys.\  {\bf 142}, 193 (1991).
}

\lref\BerkovitsBS{
  N.~Berkovits and C.~T.~Echevarria,
  ``Four point amplitude from open superstring field theory,''
Phys.\ Lett.\ B {\bf 478}, 343 (2000).
[hep-th/9912120].
}

\lref\FreedmanFR{
  D.~Z.~Freedman, S.~B.~Giddings, J.~A.~Shapiro and C.~B.~Thorn,
  ``The Nonplanar One Loop Amplitude in Witten's String Field Theory,''
Nucl.\ Phys.\ B {\bf 298}, 253 (1988).
}

\lref\ShapiroAC{
  J.~A.~Shapiro and C.~B.~Thorn,
  ``Closed String - Open String Transitions and Witten's String Field Theory,''
Phys.\ Lett.\ B {\bf 194}, 43 (1987).
}

\lref\ShapiroGQ{
  J.~A.~Shapiro and C.~B.~Thorn,
  ``{BRST} Invariant Transitions Between Closed and Open Strings,''
Phys.\ Rev.\ D {\bf 36}, 432 (1987).
}

\lref\WittenCC{
  E.~Witten,
  ``Noncommutative Geometry and String Field Theory,''
Nucl.\ Phys.\ B {\bf 268}, 253 (1986).
}

\lref\ZwiebachCS{
  B.~Zwiebach,
  ``Closed string field theory: An Introduction,''
[hep-th/9305026].
}

\lref\FradkinQD{
  E.~S.~Fradkin and A.~A.~Tseytlin,
  ``Nonlinear Electrodynamics from Quantized Strings,''
Phys.\ Lett.\ B {\bf 163}, 123 (1985).
}

\lref\TseytlinCSA{
  A.~A.~Tseytlin,
  ``On nonAbelian generalization of Born-Infeld action in string theory,''
Nucl.\ Phys.\ B {\bf 501}, 41 (1997).
[hep-th/9701125].
}

\lref\GorbonosUC{
  D.~Gorbonos and B.~Kol,
  ``A Dialogue of multipoles: Matched asymptotic expansion for caged black holes,''
JHEP {\bf 0406}, 053 (2004).
[hep-th/0406002].
}

\lref\LambertZR{
  N.~D.~Lambert, H.~Liu and J.~M.~Maldacena,
  ``Closed strings from decaying D-branes,''
JHEP {\bf 0703}, 014 (2007).
[hep-th/0303139].
}

\lref\GaiottoRM{
  D.~Gaiotto, N.~Itzhaki and L.~Rastelli,
  ``Closed strings as imaginary D-branes,''
Nucl.\ Phys.\ B {\bf 688}, 70 (2004).
[hep-th/0304192].
}

\lref\ChenFP{
  B.~Chen, M.~Li and F.~L.~Lin,
  ``Gravitational radiation of rolling tachyon,''
JHEP {\bf 0211}, 050 (2002).
[hep-th/0209222].
}

\lref\JackiwNM{
  R.~Jackiw, V.~P.~Nair, S.~Y.~Pi and A.~P.~Polychronakos,
  ``Perfect fluid theory and its extensions,''
J.\ Phys.\ A {\bf 37}, R327 (2004).
[hep-ph/0407101].
}

\lref\CarterWV{
  B.~Carter,
  ``Essentials of classical brane dynamics,''
Int.\ J.\ Theor.\ Phys.\  {\bf 40}, 2099 (2001).
[gr-qc/0012036].
}

\lref\vanHolten{
   J.W.~van Holten,
   ``Relativistic fluid dynamics,''
   {http://www.nikhef.nl/~t32/relhyd.pdf}
}

\lref\BreckenridgeTT{
  J.~C.~Breckenridge, G.~Michaud and R.~C.~Myers,
  ``More D-brane bound states,''
Phys.\ Rev.\ D {\bf 55}, 6438 (1997).
[hep-th/9611174].
}

\lref\CostaZD{
  M.~S.~Costa and G.~Papadopoulos,
  ``Superstring dualities and p-brane bound states,''
Nucl.\ Phys.\ B {\bf 510}, 217 (1998).
[hep-th/9612204].
}

\lref\HarmarkWV{
  T.~Harmark,
  ``Supergravity and space-time noncommutative open string theory,''
JHEP {\bf 0007}, 043 (2000).
[hep-th/0006023].
}

\lref\CaldarelliPZ{
  M.~M.~Caldarelli, R.~Emparan and M.~J.~Rodriguez,
  ``Black Rings in (Anti)-deSitter space,''
JHEP {\bf 0811}, 011 (2008).
[arXiv:0806.1954 [hep-th]].
}

\lref\Eckart{
   C.~Eckart,
   ``The Thermodynamics of Irreversible Processes. III. Relativistic Theory of the Simple Fluid,'' 
   Phys. Rev., 58, 919-924, (1940).
}

\lref\LF{
   L.D.~Landau and E.M.~Lifshitz, 
   ``Fluid Mechanics, vol. 6 of Course of Theoretical Physics'', 
   (Pergamon; Addison-Wesley, London, U.K.; Reading, U.S.A., 1959)
}

\lref\StewartI{
   J.M.~Stewart,
   ``On transient relativistic thermodynamics and kinetic theory'', 
   Proc. R. Soc. London, Ser. A, 357, 59-75, (1977)
}

\lref\StewartII{
   W.~Israel and J.M.~Stewart, 
   ``On transient relativistic thermodynamics and kinetic theory.II'', 
   Proc. R. Soc. London, Ser. A, 365, 43-52, (1979)
}

\lref\StewartIII{
   W.~Israel, and J.M.~Stewart, 
   ``Transient Relativistic Thermodynamics and Kinetic Theory'',
Ann. Phys. (N.Y.), 118, 341-372, (1979)
}

\lref\ArmasRVA{
  J.~Armas and T.~Harmark,
  ``Constraints on the effective fluid theory of stationary branes,''
JHEP {\bf 1410}, 63 (2014).
[arXiv:1406.7813 [hep-th]].
}

\lref\ArmasGOA{
  J.~Armas,
  ``(Non)-Dissipative Hydrodynamics on Embedded Surfaces,''
JHEP {\bf 1409}, 047 (2014).
[arXiv:1312.0597 [hep-th]].
}

\lref\ArmasHSA{
  J.~Armas,
  ``How Fluids Bend: the Elastic Expansion for Higher-Dimensional Black Holes,''
JHEP {\bf 1309}, 073 (2013).
[arXiv:1304.7773 [hep-th]].
}

\lref\ArmasJG{
  J.~Armas and N.~A.~Obers,
  ``Relativistic Elasticity of Stationary Fluid Branes,''
Phys.\ Rev.\ D {\bf 87}, no. 4, 044058 (2013).
[arXiv:1210.5197 [hep-th]].
}

\lref\ArmasNEA{
  J.~Armas and M.~Blau,
  ``New Geometries for Black Hole Horizons,''
JHEP {\bf 07}, 048 (2015).
[arXiv:1504.01393 [hep-th]].
}

\lref\DiVecchiaVM{
  P.~Di Vecchia, A.~Liccardo, R.~Marotta and F.~Pezzella,
  ``On the gauge/gravity correspondence and the open/closed string duality,''
Int.\ J.\ Mod.\ Phys.\ A {\bf 20}, 4699 (2005).
[hep-th/0503156].
}

\lref\DiVecchiaAE{
  P.~Di Vecchia, A.~Liccardo, R.~Marotta and F.~Pezzella,
  ``Gauge / gravity correspondence from open / closed string duality,''
JHEP {\bf 0306}, 007 (2003).
[hep-th/0305061].
}

\lref\KiritsisTX{
  E.~Kiritsis,
  ``Supergravity, D-brane probes and thermal superYang-Mills: A Comparison,''
JHEP {\bf 9910}, 010 (1999).
[hep-th/9906206].
}

\lref\DouglasYP{
  M.~R.~Douglas, D.~N.~Kabat, P.~Pouliot and S.~H.~Shenker,
  ``D-branes and short distances in string theory,''
Nucl.\ Phys.\ B {\bf 485}, 85 (1997).
[hep-th/9608024].
}

\lref\GubserKV{
  S.~S.~Gubser, A.~Hashimoto, I.~R.~Klebanov and M.~Krasnitz,
  ``Scalar absorption and the breaking of the world volume conformal invariance,''
Nucl.\ Phys.\ B {\bf 526}, 393 (1998).
[hep-th/9803023].
}

\lref\deAlwisMI{
  S.~P.~de Alwis,
  ``Supergravity the DBI action and black hole physics,''
Phys.\ Lett.\ B {\bf 435}, 31 (1998).
[hep-th/9804019].
}

\lref\GubserIU{
  S.~S.~Gubser and A.~Hashimoto,
  ``Exact absorption probabilities for the D3-brane,''
Commun.\ Math.\ Phys.\  {\bf 203}, 325 (1999).
[hep-th/9805140].
}

\lref\IntriligatorAI{
  K.~A.~Intriligator,
  ``Maximally supersymmetric RG flows and AdS duality,''
Nucl.\ Phys.\ B {\bf 580}, 99 (2000).
[hep-th/9909082].
}

\lref\KhouryHZ{
  J.~Khoury and H.~L.~Verlinde,
  ``On open - closed string duality,''
Adv.\ Theor.\ Math.\ Phys.\  {\bf 3}, 1893 (1999).
[hep-th/0001056].
}

\lref\DanielssonZE{
  U.~H.~Danielsson, A.~Guijosa, M.~Kruczenski and B.~Sundborg,
  ``D3-brane holography,''
JHEP {\bf 0005}, 028 (2000).
[hep-th/0004187].
}

\lref\PolyakovAF{
  A.~M.~Polyakov,
  ``Gauge fields and space-time,''
Int.\ J.\ Mod.\ Phys.\ A {\bf 17S1}, 119 (2002).
[hep-th/0110196].
}

\lref\AmadorJU{
  X.~Amador, E.~Caceres, H.~Garcia-Compean and A.~Guijosa,
  ``Conifold holography,''
JHEP {\bf 0306}, 049 (2003).
[hep-th/0305257].
}

\lref\HashimotoSM{
  A.~Hashimoto and N.~Itzhaki,
  ``Observables of string field theory,''
JHEP {\bf 0201}, 028 (2002).
[hep-th/0111092].
}

\lref\AlishahihaAS{
  M.~Alishahiha and M.~R.~Garousi,
  ``Gauge invariant operators and closed string scattering in open string field theory,''
Phys.\ Lett.\ B {\bf 536}, 129 (2002).
[hep-th/0201249].
}

\lref\DrukkerCT{
  N.~Drukker,
  ``Closed string amplitudes from gauge fixed string field theory,''
Phys.\ Rev.\ D {\bf 67}, 126004 (2003).
[hep-th/0207266].
}

\lref\GiataganasMLA{
  D.~Giataganas and K.~Goldstein,
  ``Tension of Confining Strings at Low Temperature,''
JHEP {\bf 1502}, 123 (2015).
[arXiv:1411.4995 [hep-th]].
}



\rightline{CCQCN-2015-108}
\rightline{CCTP-2015-21} 
\vskip 10pt
\Title{}
{\vbox{\centerline{Open/closed string duality and relativistic fluids}
}}
\centerline{Vasilis Niarchos}
\bigskip
\centerline{{\it Crete Center for Theoretical Physics}}
\centerline{{\it \& Crete Center for Quantum Complexity and Nanotechnology}}
\centerline{\it Department of Physics, University of Crete, 71303, Greece}
\bigskip\bigskip\smallskip
\centerline{\bf Abstract}
\bigskip

\noindent 
We propose an open/closed string duality in general backgrounds extending previous ideas
about open string completeness by Ashoke Sen. Our proposal sets up a general version of holography 
that works in gravity as a tomographic principle. We argue, in particular, that previous expectations of a
supergravity/Dirac-Born-Infeld (DBI) correspondence are naturally embedded in this conjecture 
and can be tested in a well-defined manner. As an example, we consider the correspondence 
between open string field theories on extremal D-brane setups in flat space in the large-$N$, 
large 't Hooft limit, and asymptotically flat solutions in ten-dimensional type II supergravity. We focus on a 
convenient long-wavelength regime, where specific effects of higher-spin open string modes can be 
traced explicitly in the dual supergravity computation. For instance, in this regime we show how 
the full abelian DBI action arises from supergravity as a straightforward reformulation of relativistic 
hydrodynamics. In the example of a $(2+1)$-dimensional open string theory this reformulation 
involves an abelian Hodge duality. We also point out how different deformations of the DBI action, 
related to higher-derivative corrections and non-abelian effects, can arise in this context as 
deformations in corresponding relativistic hydrodynamics.

\vfill
\Date{}


\listtoc
\writetoc
\writedefs

\newsec{Introduction}
\seclab\intro

The primary goal of this paper is to flesh out the possibility of a general holographic connection between
open and closed strings in generic backgrounds (including flat space). We will formulate an open/closed 
string duality based on the conjecture that open string theories are self-consistent quantum mechanical 
systems without the need to include explicitly couplings to closed strings. The idea of open string 
completeness has appeared previously in work by Ashoke Sen in the context of unstable D-branes 
\refs{\SenIV,\SenNF}, and is closely related to previous observations in studies of open string field theory.
We will review Sen's proposal in section \completeness.

In the context of large-$N$ type II open string theories in flat space we postulate a conjecture that opens the
road to an extension of standard examples of the AdS/CFT correspondence beyond the low-energy/near-horizon 
limit. We emphasize that this extension is conceptually distinct from previous attempts to formulate holography 
in flat space by seeking the rules of a suitable holographic dictionary on an asymptotic boundary (see 
e.g.\ attempts \refs{\ArcioniXX\ArcioniTD\DappiaggiCI-\BarnichEB} based on the BMS group 
\refs{\BondiPX,\SachsZZA}, or other attempts like \refs{\deBoerVF,\SolodukhinGS}). 


Similar ideas based on open/closed string duality have been proposed by several authors in the past
in the context of the AdS/CFT correspondence and extensions of the correspondence beyond AdS. 
A characteristic (but not exhaustive) sample of previous works that are closely related to our proposal include 
\refs{\DouglasYP\GubserKV\deAlwisMI\GubserIU\IntriligatorAI\KhouryHZ\DanielssonZE\PolyakovAF
\DiVecchiaAE\AmadorJU-\DiVecchiaVM}. We propose that Sen's completeness conjecture helps
streamline and extend certain aspects of previous discussions.

\bigskip

Testing a duality between open and closed strings in critical higher-dimensional spacetimes 
is admittedly a complicated task. We will attempt to uncover favorable evidence for a precise dictionary
in a convenient long-wavelength regime in the large-$N$, large 't Hooft limit in a special subsector of
the full dynamics (related to abelian singleton dynamics).
Unlike the low-energy/near-horizon limit, in the long-wavelength regime of interest it will be possible 
to keep explicitly effects from the whole open string tower on the open string side. 
At the same time, the standard large-$N$, large 't Hooft limit facilitates a tractable description on the closed string side 
in terms of classical supergravity.
 
We can summarize the main elements of the evidence we provide in the following way. 
Taking the traditional path of the 90s that led to the AdS/CFT correspondence, we consider the properties of 
extremal (multi-charge) $p$-brane solutions in supergravity. In a specific derivative expansion scheme of the gravitational
equations of motion we argue that the study of the long-wavelength 
perturbations of $p$-brane solutions leads naturally to an effective $(p+1)$-dimensional screen 
outside the near-horizon region where an abelian effective action can be formulated. 
Following previous discussions in the context of the blackfold formalism \refs{\EmparanCS,\EmparanAT} 
we postulate that there is a one-to-one correspondence between the solutions of the equations of motion 
of this effective action and a certain class of regular solutions of the full-fledged gravitational equations. 
For extremal solutions in flat space we prove that the action on the gravitational effective screen 
is identical to the abelian Dirac-Born-Infeld (DBI) action. We argue that the latter is the abelian part of the Wilsonian 
effective action of the holographically dual large-$N$ open string theory making a precise connection between 
gravity and open strings. We discuss how non-abelian effects are incorporated in this picture.

\bigskip

Our general conjecture for a duality between open and closed strings is formulated in section \OCduality. 
In the same section we describe some of the anticipated features of the Wilsonian effective action of open strings, 
the specifics of the long-wavelength expansions of interest and the implementation of these expansions on 
the gravitational side. Our main task in this section is to collect and organize the accumulating observations 
over the years into a coherent story under a single framework. For many of the underlying technical details of the topics 
that enter this subject we direct the reader to the appropriate references.

One of the main technical tasks of this paper is to verify that the abelian theory on the gravitational 
effective screen coincides with the one expected on the open string side. In complete analogy to the 
fluid/gravity correspondence in AdS/CFT \BhattacharyyaJC, we find that the effective theory that 
emerges naturally in gravity is formulated in the form of relativistic hydrodynamics. Consequently, 
open/closed string duality in this context requires a connection between relativistic hydrodynamics 
and open string effective actions. At zero temperature and finite chemical potential, we show that 
there is indeed such a direct connection involving the DBI action. Specifically, in section \dbisugra\ we recover 
the abelian DBI action from an (anisotropic) hydrodynamic theory of fluids on dynamic elastic hypersurfaces. 
Sections \link, \fluidreview\ and \duality\ prepare the connection between hydrodynamics and gauge theory from a purely 
hydrodynamic point of view (that as far as we know is novel). 
  
The emergence of the abelian DBI theory in the context of extremal $p$-brane solutions in supergravity has a long 
history and its relation to a putative open/closed string duality has been widely anticipated. Sometimes this relation is 
referred to in the literature as the supergravity/DBI correspondence. From this perspective 
two of the main new contributions in this paper are: 
\item{$(i)$} We propose that the supergravity/DBI correspondence can be made into an algorithmic map within the general 
formalism of blackfolds in supergravity (extending the proposal of our recent work \NiarchosMAA). 
In the present paper we provide an important part of this map: the explicit relation between the fluid dynamical 
variables of the gravitational long-wavelength description and the gauge-theoretic degrees of freedom of the open string description, and the precise relation between the equations of motion that both degrees of freedom obey at extremality. 
To the best of our knowledge, the details of this relation have not been exhibited before. In fact, the key role that 
long-wavelength expansions in supergravity play in this connection has not been appreciated. Previous  
investigations have focused, almost unanimously, on exact (mostly supersymmetric) supergravity solutions, where
the connection with abelian DBI (see Refs.\ \refs{\EmparanUX\LuninMJ-\LuninTF} for some examples)
and DBI-related structures (e.g.\ calibrations \GauntlettDI) has been noticed more on the level of 
observation and less on the level of a systematic exploration. 
\item{~}
As we pointed out in \NiarchosMAA\ there is a related old approach (first applied to string theory in \CallanKY) that 
identifies the abelian part of the brane degrees of freedom in supergravity as collective coordinates associated to large gauge
transformations (for a review see \Simon). A notable improvement of the blackfold approach is that it 
encodes rather easily the full non-linear nature of the DBI action, which is hard to achieve with the techniques of
\CallanKY.
\item{$(ii)$} The AdS/CFT correspondence is embedded naturally in the big picture that we postulate 
by taking the standard low-energy/near-horizon limit of Maldacena \MaldacenaRE. 
Several authors in the past have pointed out the importance of the 
singleton degrees of freedom for physics outside the near-horizon throat. We re-emphasize the key role played
by singleton degrees of freedom and point out that the abelian effective actions that we discuss are singleton
effective actions embedded naturally within the full non-abelian Wilsonian effective action of a dual open string
theory. The discussion at this point is closely related to the pre-AdS/CFT considerations of Ref.\ \GibbonsSV. 
We sketch how non-abelian effects can be incorporated in the singleton descriptions 
by integrating out interactions between abelian and non-abelian degrees of freedom.

Finally, the emergence of hydrodynamics in the above story is interesting for independent reasons. 
For example, there has been renewed interest in recent investigations, e.g.\ \refs{\DubovskySJ\HaehlPJA-\deBoerIJA}, 
in potential reformulations of fluid dynamics in terms of a Lagrangian variational principle. 
Our results provide an explicit illustration of an extremal hydrodynamic system where the passage 
to an action principle is facilitated by a convenient change of variables to a new set of degrees of freedom. 
The latter are clearly the degrees of freedom favored in the Wilsonian effective description of the underlying 
(open string) microscopics. 

The interplay between hydrodynamics and open string theory holds the promise of interesting lessons about both 
frameworks. For example, from the gravitationally-derived hydrodynamics we learn about various deformations 
of the abelian DBI action induced by large-$N$ non-abelian effects. Moreover, the exact gravitational solutions 
provide an efficient resummation of all the DBI higher-derivative corrections. 
Conversely, by studying higher-derivative corrections of the DBI action in open string theory we can make 
predictions using open/closed string duality about higher-derivative corrections of hydrodynamics that 
have not yet been computed in supergravity. A preliminary study of this aspect appears in section \higherderiv.

\newsec{Sen's open string completeness revisited}
\seclab\completeness

Before we go into the specifics of our proposal, let us briefly recall a closely related circle of ideas 
about open/closed string duality put forward by Sen in Ref.\ \SenIV\ in the context of unstable D-branes in string 
theory.

Unstable D-branes exhibit a rapid time-dependent decay into closed strings \refs{\LambertZR\GaiottoRM-\ChenFP}. 
During this process closed strings with typical energies of order $1/g_s$ are copiously produced until the end of the 
process where the D-brane (and the open strings on it) completely disappear. Yet, it was observed at tree-level 
(in the limit where $g_s \to 0$) that the open string description of this process in terms of a rolling tachyon manages 
to reproduce correctly many of the features of this process {\it at all times} without the need to include explicit 
open/closed string couplings. A review of this evidence can be found in \SenNF. This led Sen to conjecture in Ref.\ \SenIV\ 
that

\vskip 0.1cm
\noindent
{\it there is a quantum open string field theory (OSFT) that describes the full dynamics of the unstable D-brane without
an explicit coupling to closed strings.}

\vskip 0.1cm
This statement is consistent with independent studies of open string field theory demonstrating formally that the 
perturbative expansion of OSFT around the maximum of the tachyon potential is complete 
\refs{\GiddingsWP\ZwiebachAZ-\BerkovitsBS}, and that open string amplitudes have the correct poles 
associated to intermediate closed string states \refs{\FreedmanFR\ShapiroAC-\ShapiroGQ}
(see also \refs{\HashimotoSM\AlishahihaAS-\DrukkerCT}). The validity of
the above conjecture was further tested by Sen in the context of unstable D0-branes in two-dimensional 
non-critical string theory using the correspondence with double-scaled matrix models.

The assumption that the OSFT on a D-brane setup is a self-consistent quantum mechanical system, implies that
OSFT contains complete information about the closed strings produced by the D-brane, and therefore 
suggests that states in closed string theory can be described holographically in a non-gravitational language. 
This does {\it not} imply that a given OSFT can describe any closed string state. It can only describe those states produced 
by the decaying D-brane. This introduces an interesting way to think about closed string theory and 
gravitational dynamics, where we split closed string solutions into separate, quantum mechanically self-consistent, 
superselection sectors. In this manner, Sen's open/closed string duality works as {\it tomography}, where 
different OSFTs slice through different subsectors of the vast configuration space of gravity and closed strings.

There are several apparent differences between this version of open/closed string duality and the more familiar 
gauge/gravity dualities in the AdS/CFT correspondence. For instance, in the example of Sen there is no 
large-$N$ limit, and closed strings are produced via a time-dependent process.
The correspondence is expected to work even for a single unstable D-brane with an abelian OSFT.
Instead, in the large-$N$ limit of standard AdS/CFT examples closed strings are produced by heavy, 
typically static and stable, D-brane configurations. Finally, in AdS/CFT there is a clear holographic screen 
(boundary) where the dual non-gravitational theory is naturally envisioned. No such screen is visible in Sen's proposal.

In the next section, we conjecture a general framework based on Sen's ideas that attempts to 
bridge the gap of these apparent disparities.

\newsec{Open/closed string duality and flat space holography as a special case}
\seclab\OCduality

A universally expected feature of any type of duality between two theories {\bf A} and {\bf B} 
(including holographic dualities) is an equality between their respective effective actions
\eqn\generalaa{
\SS_{\bf A}\left[ \Phi_{\bf A}, \JJ_{\bf A} \right]  = \SS_{\bf B} \left[ \Phi_{\bf B}, \JJ_{\bf B} \right]
}
under a specific map between the collection of vacuum expectation values $\Phi_{\bf A}$ and $\Phi_{\bf B}$ that label
the vacuum state on both sides, and the generic sources $\JJ_{\bf A}$ and $\JJ_{\bf B}$ that represent 
deformations of the two theories. In quantum mechanical systems with standard Lagrangian formulations the 
(Wilsonian) effective action is defined formally by a path integral over field configurations $\phi$ of the form
\eqn\generalab{
\SS = - \log Z = - \log \left(   \int [d\phi] \, e^{-\int \LL[\phi]} \right)
~.}

For example, in the supergravity regime of the standard AdS/CFT correspondence theory {\bf A} is a large-$N$ quantum
gauge theory, and theory {\bf B} is a classical supergravity theory in asymptotically AdS spacetimes.
The quantities $\Phi_{\bf A}$ are vacuum expectation values (vevs) of gauge-invariant operators $\OO_{\bf A}$,
i.e.\ $\Phi_{\bf A}=\langle \OO_{\bf A}\rangle $, and $\JJ_{\bf A}$
are external sources for the same operators. The quantities $\Phi_{\bf B}$ label the classical profiles of the supergravity 
fields in gravitational solutions with specified asymptotics $\JJ_{\bf B}$. The standard holographic dictionary in 
AdS/CFT explains how one translates the pairs $(\Phi_{\bf A},\JJ_{\bf A})$ to the pairs $(\Phi_{\bf B}, \JJ_{\bf B})$.

\subsec{The proposed conjecture}
\subseclab\general

We would like to conjecture a more general holographic duality with the following ingredients. 
Theory {\bf A} is a (non-abelian) open string field theory on a stack of D-branes  
embedded in a specified closed string background $\Psi$.\foot{The specification of a closed string background in the form
of a given two-dimensional worldsheet conformal field theory is a standard basic ingredient in all known formulations
of open string field theory including Witten's cubic open string field theory \WittenCC\ and subsequent extensions.} 
Theory {\bf B} is a closed string field theory\foot{A workable formulation of closed string field theory is a notoriously 
hard technical problem, see \ZwiebachCS\ for a review of early attempts. Here, we refer to closed string field theory 
as a putative quantum mechanically consistent framework for closed strings, that we assume to exist beyond the 
standard first-quantized perturbative formulations.} 
that has $\Psi$ as a vacuum state and theory {\bf A} as one of its allowed 
open string sectors. We add the following requirement: the D-brane configuration should guarantee the existence of 
an asymptotic region where the D-brane backreaction on closed strings is negligible and the closed string fields 
asymptote to $\Psi$. In general, this requirement puts constraints on the co-dimension of the D-brane configuration, 
and introduces two standard features of holography: a holographic radial direction, and an asymptotic region that 
does not change under normalizable deformations.

In this context $\Phi_{\bf A}$ are gauge-invariant vevs of open string fields and $\JJ_{\bf A}$ are the open-closed 
string couplings induced by placing the open string theory on the background $\Psi$. On the other hand, 
$\Phi_{\bf B}$ label the profiles of closed string field vevs in states with asymptotics defined by $\Psi$, and 
$\JJ_{\bf B}$ coincides with the asymptotic closed string state $\Psi$.

As a concrete example, let us consider the open string theory that resides on a D-brane setup in flat space in perturbative 
ten-dimensional type II string theory (e.g.\ open string theory on a stack of D3-branes in flat space). 
In open-closed string theory we can compute bulk-boundary couplings by 
evaluating the 2-point functions of one open string going to one closed string. From the worldsheet point of view, 
at leading order in $g_s$, this involves the disc 2-point function $\langle \VV_{open} \VV_{closed}\rangle_{disc}$,
where $\VV_{open}$ is a vertex operator inserted at the boundary of the disc and $\VV_{closed}$ is a vertex operator
inserted at the center of the disc. Hence, in a small deformation of flat space with non-vanishing profile $\JJ_{\bf B}$ 
of the closed string fields these couplings dictate uniquely how open string fields are sourced, i.e.\ they determine the external 
couplings that we called collectively $\JJ_{\bf A}$. 

Consequently, given a collection $\JJ_{\bf B}$ we can, in principle, compute independently two separate quantities: 
$(i)$ the quantum effective action $\SS_{\bf A}$ of the open string theory (without explicit coupling to closed strings) 
in the presence of the sources $\JJ_{\bf A}$ (deduced from the profile of $\JJ_{\bf B}$), in an open string vacuum 
labeled by $\Phi_{\bf A}$, and $(ii)$ the quantum effective action $\SS_{\bf B}$ of closed string theory 
(without any coupling to open strings) in a vacuum where the closed string fields asymptote to $\JJ_{\bf B}$. 
We conjecture that there is an one-to-one map between the open and the closed string vacua in points $(i)$ and 
$(ii)$ such that $\SS_{\bf A}$ and $\SS_{\bf B}$ are identical.

In the following subsections we will explore the validity of the above conjecture in a technically convenient large-$N$, 
long-wavelength regime, where the description on the closed string side reduces to standard classical supergravity, 
but the open string side is stringy and we can keep track of explicit open string effects at all orders in $\alpha'$.

\subsec{Quantum open string effective actions}
\subseclab\effective

On the open string side we consider the Wilsonian effective action $\SS \equiv \SS_{\bf A}[\Phi_{\bf A},\JJ_{\bf A}]$
of a non-abelian open string theory on a stack of D-branes. For concreteness and simplicity of the presentation we 
will concentrate on the special case of $N$ coincident D$p$-branes $(p<6)$ in flat space in perturbative ten-dimensional type II 
string theory. The dynamics of open strings in such setups is described by a non-abelian open string field theory where
the open string fields are fields in the adjoint representation of $U(N)$. At low energies this theory reduces to a $U(N)$ 
supersymmetric gauge theory. Besides the rank of the gauge group $N$, the other free parameter that characterizes 
the open string theory in this setup is the string coupling $g_s$, which is part of the sources $\JJ_{\bf A}$ 
(determined by the closed string background). 

For reasons that are clear already in standard AdS/CFT discussions (and will be repeated in the next subsection) 
it is convenient to consider the large-$N$ 't Hooft limit where $N\gg 1$, $g_s\ll 1$ and $\lambda = g_s N$ is a fixed 
tunable parameter.

The $U(N)$ open string field theory on the D-branes is a complicated quantum mechanical system 
with an infinite tower of interacting open string modes. The Wilsonian effective action in a generic open string vacuum 
is computed formally from a complicated path integral in string field theory of the general abstract form \generalab. In a weak 
't Hooft coupling expansion (valid when $\lambda \ll 1$) this action, which is a function of the open string field vevs 
$\Phi_{\bf A}$, is expressed by a series of the form
\eqn\openaa{\eqalign{
\SS \left[ \Phi_{\bf A}; \lambda, N\right]
=& \sum_{n=0}^\infty \lambda^{-1+n} \SS_n [\Phi_{\bf A}; N] 
\cr
=& \sum_{g=0, h=1}^\infty N^h g_s^{2g-2+h} \SS_{g,h}[\Phi_{\bf A}]
= \sum_{g=0, h=1}^\infty N^{2-2g} \lambda^{2g-2+h} \SS_{g,h}[\Phi_{\bf A}]
~.}}
From the worldsheet point of view $\SS_{g,h} \left[ \Phi_{\bf A}; N\right]$ is an off-shell partition function on a 
Riemann surface with $g$ handles and $h$ holes \FradkinQD. At low-energies the corrections in this expansion 
arise from perturbative loop diagrams in quantum gauge theory. Non-planar diagrams contribute terms with 
subleading $1/N$-dependence in the large-$N$ limit.

In the opposite regime, at strong 't Hooft coupling, we expect that the more appropriate description of the effective action 
is not in terms of the `elementary' open string vevs $\Phi_{\bf A}$, but in terms of vevs of gauge-invariant composites, 
let us call them $\tilde \Phi_{\bf A}$. Accordingly, we expect a different expansion in inverse powers of $\lambda$ of the form
\eqn\openab{
\SS \left[ \tilde \Phi_{\bf A}; \lambda, N\right]
= \sum_{n=0}^\infty \lambda^{\alpha-n} \tilde \SS_n[\tilde \Phi_{\bf A}; N]
~,}
where $\alpha$ is an appropriate constant. Soon we will suggest that the value of $\alpha$ is $-1$.

In an effort to be more concrete about the evaluation of these expansions, let us 
concentrate for starters on the first term, $\SS_0$, of the weak coupling expansion. 
This term is essentially classical, and as we mentioned, it can be computed in string perturbation 
theory from the disc partition function $(g=0, h=1)$. Although the exact general result of this computation is not known, 
in the past it has been extremely useful to think about $\SS_0$ in a long-wavelength expansion of the vevs $\Phi_{\bf A}$ 
around a suitably symmetric vacuum.

In this approach an open string vacuum is characterized by the vevs of the low-lying massless open string fields, that 
include among other things the non-abelian gauge field $A_a$ and transverse scalars $X^\perp$. The indices 
$a=0,1,\ldots, p$ are D-brane worldvolume indices and $\perp$ summarizes collectively the background spacetime indices 
perpendicular to the brane. The massive open string modes are integrated out and their quantum effects are incorporated in 
higher order interactions between the massless fields. 

The vacuum around which the long-wavelength expansion is set up must be an exact open string vacuum.
In the abelian case it is known that the vacua with arbitrary constant transverse velocities $\d_a X^\perp$ and constant gauge 
field strength $F_{ab}$ are such vacua. The long-wavelength expansion is an expansion in derivatives of these 
quantities\foot{$\SS_0$ also includes terms for the action of the superpartner fermions that will be kept implicit in the 
following discussion.}
\eqn\openaca{
\SS_0 [A_a, X^\perp ; N] = S_{0}[F_{ab}, \d_a X^\perp] + S_{higher-order}[\d^n F , \d^m X]
~.}
The leading term $S_0$ on the RHS is exact on the field strength $F$ and the velocities $\d X^\perp$ at all orders in $\alpha'$ and
$S_{higher-order}$ is a perturbative expansion in higher derivatives of $F$ and $\d X^\perp$. A well defined computation
in open string theory determines $S_0$ as the abelian DBI action (see \TseytlinDJ\ for a review and list of original 
references)
\eqn\openad{
S_{DBI} = \TT_p \int d^{p+1}x\, \sqrt{-\det (\eta_{\mu\nu} \d_a X^\mu \d_b X^\nu + 2\pi \alpha'  F_{ab})}
~,}
where $\TT_p = \frac{1}{g_s \sqrt{\alpha'}} \frac{1}{(2\pi \sqrt \alpha')^p}$ is the D$p$-brane tension, and $\mu,\nu = 0,1,\ldots, 9$
are indices for the flat spacetime background with Minkowski metric $\eta_{\mu\nu}$. 

The non-abelian case is richer and comparatively less understood. It is natural to look for a non-abelian extension 
of the above expansion as an expansion in small gauge-covariant derivatives, $D_a$, of the velocities $D_a X^\perp$ and gauge
field strength $F_{ab}$
\eqn\openacb{
\SS_0 [A_a, X^\perp ; N] = S_{0}[F_{ab}, D_a X^\perp] + S_{higher-order}[D^n F , D^m X]
~.}
However, it is already apparent from the identity 
\eqn\openacc{
[D_a,D_b]F_{cd} = [F_{ab},F_{cd}]
}
that the leading term $S_0$ is now ambiguous. Moreover, because of \openacc\ an expansion in small covariant derivatives 
will be also an expansion around commuting field strengths. Isolating the symmetric covariant derivatives in 
$S_{higher-order}$ and using \openacc\ $S_0$ comprises of two pieces ---one with $F$ commutators and one 
without.\foot{Terms that involve the covariant velocities $D_a X^\perp$ can be decided by T-duality from the 
ten-dimensional open string effective action.} Ref.\ \TseytlinCSA\ proposed that the part of $S_0$ without $F$ 
commutators is a non-abelian version of the DBI action \openad\ with a symmetric trace prescription. 
For a nice summary of the problems and progress on the non-abelian DBI action we refer the reader to \MyersBW.

Despite the above-mentioned technical difficulties, there are a few robust features expected from the putative 
non-abelian $S_0$, for gauge group $U(N)$, that are useful to highlight for later purposes.

When expanded in a series of powers in $F$ and $D X^\perp$, the leading term in $S_0$ is quadratic. In the quadratic 
interactions the vector $U(1)$ and $SU(N)$ parts completely decouple. The $SU(N)$ part includes the 
$(p+1)$-dimensional non-abelian super-Yang-Mills action. At higher orders in this power series the $U(1)$ and $SU(N)$ 
degrees of freedom are coupled by higher-dimension interactions. From this point of view we can write $S_0$ schematically 
as a sum of three terms
\eqn\openae{
S_0 = S_{0,U(1)}[\Phi] + S_{0,SU(N)}[{\bf \Phi}] + S_{0,mixed}[\Phi, {\bf \Phi}]
~,}
where we have denoted compactly the abelian degrees of freedom by (a normal font) $\Phi$ and their non-abelian 
counterparts by (a bold font) $\bf \Phi$. Setting ${\bf \Phi}=0$ (by which we mean explicitly $F=0$, $D X^\perp=0$ 
for the $SU(N)$ vevs) we deduce that $S_{0,U(1)}$ is simply $N$ times the abelian DBI action \openad
\eqn\openaf{
S_{0,U(1)} = N S_{abelian\, DBI}
~.
}
$S_{0,mixed}$ involves interactions at cubic and higher order between $\Phi$ and $\bf \Phi$.

Varying separately with respect to $\Phi$ and $\bf \Phi$ to determine the abelian and non-abelian vacuum expectation 
values, we find two coupled sets of equations
\eqn\openag{\eqalign{
&\frac{\delta S_{0,U(1)}}{\delta \Phi} [\Phi]+ \frac{\delta S_{0,mixed}}{\delta \Phi} [\Phi, {\bf \Phi}] = 0 
~,
\cr
&\frac{\delta S_{0,SU(N)}}{\delta {\bf \Phi}} [{\bf \Phi}]+ \frac{\delta S_{0,mixed}}{\delta {\bf \Phi}} [\Phi, {\bf \Phi}] = 0
~,}}
where $\delta$ denotes the standard Euler-Lagrange variation. Since terms in $S_{0,mixed}$ have to be at least quadratic
in the non-abelian fields, setting ${\bf \Phi} = 0$ 
is a consistent ansatz that satisfies automatically the second equation in \openag\ and leaves 
\eqn\openah{
\frac{\delta S_{0,U(1)}}{\delta \Phi} [\Phi] = 0
}
from the first equation. 

Notice the following important property. In a state with a non-trivial abelian part that solves \openah, 
the higher-dimension interactions in $S_{0,mixed}$ are {\it not} infrared irrelevant. In other words, 
around a non-trivial abelian vacuum the non-abelian dynamics does not implicate only terms from 
$S_{0,SU(N)}$ but also terms from $S_{0,mixed}$.

We postulate that the structure \openae\ applies to the full effective action $\SS$, 
not just the classical contribution $S_0$. For example, in the regime of strong 't Hooft coupling
the vevs of gauge-invariant operators are separated naturally into vevs of the abelian fields $\tilde \Phi$, and
vevs of the non-abelian fields $\tilde {\bf \Phi}$. We expect that it is possible to express the 
equations of motion of the strong coupling effective action \openab\ in a form analogous to \openag, and that the
trivial non-abelian vevs ${\bf \Phi}=0$, or $\tilde {\bf \Phi}=0$, are a consistent ansatz for the full effective action $\SS$.
We will refer to the vacua with trivial non-abelian vevs as {\it the origin of the Coulomb branch} of the open string 
field theory.

\medskip
\noindent
{\bf Main aim of the paper:} we focus first and foremost on the dynamics of the abelian part of the 
effective action $\SS$ at the origin of the Coulomb branch in the long-wavelength approximation, 
and compare descriptions of this sector at weak and strong t' Hooft coupling.

\medskip

The first thing to notice about the abelian sector in the long-wavelength approximation is that the 
abelian vevs of $F$ and $\d X^\perp$ are automatically gauge-invariant. Hence, there is potential for a direct
relation between the weak coupling description of these degrees of freedom and their corresponding strong coupling 
description.

Restricting our attention to the origin of the Coulomb branch (${\bf \Phi}=0$), the small 't Hooft coupling expansion
\openaa\ becomes an expansion in terms of the abelian vevs $\Phi$. Naively the effective action receives corrections from 
worldsheets with an arbitrary number of handles and holes (each one being of the order $\OO(N^h g_s^{2g-2+h})$).
Nevertheless, we will soon find from a direct supergravity analysis valid at the strong 't Hooft coupling limit that at 
leading order in the long-wavelength derivative expansion the leading term in the $1/N$ expansion of \openab\ 
is identical to the leading term in the $1/N$ expansion of the weak coupling series \openaa, namely 
\eqn\openaia{
\SS_0 \big |_{leading\, 1/N, leading\ derivative, U(1)} = 
\tilde \SS_0 \big |_{leading\, 1/N, leading\, derivative, U(1)}
~.
}

This observation suggests the possibility of an even stronger relation valid at all orders of the derivative expansion
\eqn\openai{
\SS_0 \big |_{leading\, 1/N, U(1)} = 
\tilde \SS_0 \big |_{leading\, 1/N, U(1)}
~.
}
This equation implies a non-renormalization theorem of the open string effective action at the origin of the 
Coulomb branch, where the corrections from worldsheets with more than one hole $(h>1)$ are vanishing and the
action (expressed in terms of $N$ and $g_s$) has a trivial linear dependence on $N$ at all values of $\lambda$.
Moreover, \openaia\ (or the stronger \openai) suggest that the value of the constant $\alpha$ in \openab\ is $-1$.

Currently, we are not aware of a conclusive proof of this non-renormalization theorem. For the purposes of equation
\openai\ it would be sufficient to have a proof of the cancellation of the open string corrections coming
from worldsheets with $h>1$ at zero genus, $g=0$. Although such a cancellation may sound plausible in 
supersymmetric configurations notice that we observe \openaia\ for any extremal configuration irrespective
of supersymmetry. Assuming the existence of such a non-renormalization theorem, the derivation of \openai\
from supergravity would constitute a direct test of the open/closed string duality formulated in the beginning
of this section.

We note that equation \openaia\ is an integral part of the long-anticipated correspondence between solutions of the 
DBI action and extremal supergravity configurations, which has been observed experimentally in many examples in the past.
In what follows, we describe a framework where such a correspondence can be formulated in a more organized 
manner.

\subsec{Closed strings: long-wavelength expansions and effective actions in supergravity}
\subseclab\blackfolds

Having discussed some of the features on the open string side of the putative open/closed string duality of section 
\general\ we now pass to a corresponding discussion on the closed string side. We continue to focus at the leading 
order in the large-$N$, large 't Hooft coupling limit, where standard arguments in the context of the AdS/CFT 
correspondence show that the low-energy effective field theory description of closed string field theory, 
$\SS_{\bf B}$, is the ten-dimensional supergravity action. 

The trivial vacua of the open string theory on a stack of $N\gg 1$ D$p$-branes that we want to consider (as 
0th-order configurations in subsequent long-wavelength derivative expansions) are captured 
holographically on the supergravity side by extremal $p$-brane solutions with translation invariance
in the worldvolume directions. To incorporate more features, these solutions may also involve 
homogeneous fluxes sourced by lower-dimensional branes smeared along the 
$p$-brane worldvolume, e.g.\ the $p$-brane solution may be an F1-D$p$ bound state (corresponding to a planar 
stack of D$p$-branes with a constant worldvolume abelian electric field turned on), or a more complicated 
multi-charge bound state. Let us call this solution $\Phi_{\bf B}^{(0)}$. In this case the metric and all other 
supergravity fields depend non-trivially only on the radial coordinate $r$. Specific examples will be discussed below.
In what follows we want to consider configurations away from the trivial homogeneous vacuum, and to explore 
if these have a chance to map holographically to solutions of the open string field theory according to the conjecture 
of section \general.

As we mentioned near the end of the previous subsection, our primary goal is to compare the open string effective 
action $\SS_{\bf A}$ in the long-wavelength expansion to the closed string (supergravity) effective action 
$\SS_{\bf B}$ in a corresponding expansion. Hence, on the supergravity side it is natural to look for extremal 
{\it inhomogeneous} $p$-brane configurations of the form
\eqn\blackfoldsaa{
\Phi_{\bf B} = \Phi_{\bf B}(x^\mu)
}
expanded in small derivatives with respect to the $p$-brane worldvolume coordinates $\sigma^a$ ($a=0,1,\ldots,p$) 
\eqn\blackfoldsab{
\Phi_{\bf B} = \Phi_{\bf B}^{(0)}(r) + \varepsilon \Phi_{\bf B}^{(1)}(r,\sigma^a) + \varepsilon^2 \Phi_{\bf B}^{(2)}(r,\sigma^a) +\ldots
}
The dummy variable $\varepsilon$ keeps track of the number of worldvolume derivatives $\d_a$. This expansion is 
inserted into the PDEs of supergravity which are solved perturbatively order by order. 

Our open/closed string conjecture states that there is a unique on-shell map between the open string 
vevs $\tilde \Phi_{\bf A}$ and the closed string (supergravity) vevs $\Phi_{\bf B}$. Under this map the 
on-shell value of the open string effective action \openab\ (at leading order in the $1/N$ and 
$1/\lambda$ expansion, which is the regime of interest here) should equal the on-shell value of the 
supergravity action $\SS_{\bf B}$ at all orders in the long-wavelength derivative expansion.

We discussed why abelian deformations at the origin of the Coulomb branch is a computationally opportune context. 
What is the corresponding description of this sector on the supergravity side? 

The 0th-order supergravity profile $\Phi_{\bf B}^{(0)}$, which corresponds to an open string vacuum 
at the origin of the Coulomb branch, is labelled by a set of constants that parametrize the asymptotic 
charges, e.g. mass, angular momentum, brane charges. These parameters, and other parameters 
associated with the breaking of the global symmetries of the asymptotic background by the $p$-brane solution, 
can be viewed as collective coordinates of the supergravity solution. A restricted class of 
supergravity solutions can be constructed perturbatively around $\Phi_{\bf B}^{(0)}$ with a supergravity ansatz 
that promotes the collective coordinates into slowly-varying functions of the worldvolume coordinates. 
This is a special case of the general expansion \blackfoldsab. It has been argued long ago \GibbonsSV\ 
(and also more recently in \NiarchosMAA) that the above collective modes are the supergravity manifestation of the massless $U(1)_{vector}$ degrees of freedom of the dual open string theory. Consequently, we propose 
that the supergravity deformations within this sector are the holographic duals of the open string
configurations with vanishing non-abelian vevs. 

A systematic development of long-wavelength expansions of the type we have just described in general 
(super)gravity theories has been initiated in recent years in the context of the blackfold formalism starting from 
Refs.\ \refs{\EmparanCS,\EmparanAT}. We refer the reader to the existing literature for 
a more detailed technical exposition of current results in this (still developing) framework. Here it will be 
useful to highlight some of the key conceptual features of the formalism that play a role in our general discussion:

\vskip 0.2cm
\item{$(a)$} {\bf (Super)gravity expansions.}
Perturbative supergravity solutions in the blackfold approach are constructed using the method of 
Matched Asymptotic Expansions (MAEs) (see \GorbonosUC\ for an instructive application of MAEs to caged black holes). 
An exact 0th-order $p$-brane solution is perturbed in a {\it near-zone} region ($r\ll \RR$, where $\RR$ is the 
typical scale of the long-wavelength perturbation) by promoting the collective modes to slowly varying functions 
of the worldvolume coordinates. At the same time the supergravity fields are corrected appropriately
to achieve a perturbative solution of the full set of supergravity equations of motion. 
The ansatz in this region is performed in a very similar way conceptually to analogous 
constructions for AdS black branes in the fluid-gravity correspondence in AdS/CFT \BhattacharyyaJC. 
Simultaneously, an independent computation is performed in a Newtonian approximation far from the 
horizon in the {\it far-zone} region ($r\gg r_H$, where $r_H$ is the typical near-horizon radius of the 
$p$-brane solution; for extremal $p$-brane solutions this is the charge radius that appears explicitly in section \dbisugra).
A matched asymptotic expansion is performed order-by-order by matching the near-zone and far-zone 
solutions in the large overlap region $r_H \ll r \ll \RR$, whose existence is the basis of the long-wavelength expansion.

\vskip 0.2cm
\item{$(b)$} {\bf Collective mode equations.}
During this process one discovers that a part of the supergravity equations (constraint equations 
in the near-zone analysis) reduces to a system of $(p+1)$-dimensional equations for the collective coordinates. 
These lower-dimensional equations, that we call {\it blackfold, or collective mode equations}, are expressed 
naturally as conservation equations for a set of currents; these include the energy-momentum tensor $T_{ab}$, 
and $q$-form currents $J_{a_1\cdots a_q}$ related to the charges of the $p$-brane solution. In general bound state 
solutions there are several different values of $q\leq p$. These equations are naturally formulated as hydrodynamic
equations for fluids propagating on dynamical hypersurfaces.

\vskip 0.1cm
\item{~} One of the main technical computations below will be to exhibit the precise relation between these supergravity 
collective mode equations and the abelian DBI equations of motion making a part of the relation \openai\ manifest within the 
postulated open/closed string duality. 

\vskip 0.1cm
\item{~} Finally, it has been conjectured \refs{\EmparanAT} that solutions of the blackfold equations are in one-to-one 
correspondence with solutions of the full supergravity equations order-by-order in the long-wavelength expansion. 
A general proof of this conjecture (referred to as {\it the blackfold conjecture} below) is not
available at the moment. However, a proof in special cases at first order in the derivative expansion has appeared in 
\refs{\CampsBR,\CampsHW,\GathQYA}.

\vskip 0.2cm
\item{$(c)$} {\bf Dynamical holographic screen.}
Similar to the AdS/CFT correspondence, where the dual theory is naturally thought of as a theory 
residing on the asymptotic boundary of AdS, in the present case the dual open string theory is naturally thought 
of as a theory residing on a D-brane stack embedded in the given asymptotic background. In the main examples 
of this paper the asymptotic background is flat space. This feature is emerging almost automatically from gravity 
in the long-wavelength blackfold expansions. The currents whose conservation defines a set of lower-dimensional 
dynamical equations in supergravity (collective mode equations) are computed in the overlap region, which can 
be viewed as the asymptote of the near-zone region deep inside the asymptotically flat far-zone region. 
Hence, in the long-wavelength approximation we discover naturally within supergravity the emergence of 
a dynamical lower-dimensional holographic screen embedded in the asymptotic background.

\vskip 0.2cm
\item{$(d)$} {\bf Higher-derivative corrections.}
The order-by-order solution of the gravitational PDEs results to a perturbative higher-derivative modification of the 
blackfold equations in a fixed background. In extremal setups we postulate that such derivative corrections are in direct 
correspondence with the derivative corrections to the abelian DBI action which can be computed from
standard calculations on the disc worldsheet of the dual open string theory. This postulate assumes
the validity of the non-renormalization relation \openai. A preliminary discussion of 
this correspondence appears in section \higherderiv\ below.

\vskip 0.1cm
\item{~} It is also interesting to note in this context that we can interpret an exact inhomogeneous $p$-brane solution 
in gravity (with flat space asymptotics) as the dual of a non-perturbative resummation of the derivative expansion 
of the open string effective action including all stringy effects in the leading order in the $1/N$ and $1/\lambda$
expansions.

\vskip 0.2cm
\item{$(e)$} {\bf Low-energy/near-horizon limits.} 
The low-energy (small field-strength) limit on the open string side corresponds to the near-horizon limit of the 
(deformed) supergravity solutions. As we discussed in section \effective, around the trivial undeformed vacuum 
the low-energy limit on the open string side results to a decoupling of the $SU(N)$ and $U(1)$
sectors. The $SU(N)$ part is strongly interacting and has a dual AdS/CFT description in terms of gravity 
in the near-horizon region of the homogeneous 0th-order supergravity solution. 
The $U(1)$ part also has a well-known description in the near-horizon limit. It corresponds to singleton degrees of 
freedom that are topological in the bulk and are fully supported on the boundary. From this point of view, it is natural to think 
of the blackfold effective field theory that we formulate as a singleton effective field theory \refs{\GibbonsSV,\NiarchosMAA}.

\vskip 0.1cm
\item{~} In accordance with the discussion in section \effective, the near-horizon limit around a {\it deformed solution} does not
have to lead to a complete infrared decoupling of the singleton degrees of freedom. In fact, in some cases a deformed brane
solution may not even have a single near-horizon limit as BIon-type solutions in supergravity exhibit 
\refs{\LuninMJ,\GrignaniXM,\NiarchosPN}.

\vskip 0.2cm
\item{$(f)$} {\bf Non-abelian effects.}
The holographic encoding of the strongly interacting non-abelian degrees of freedom of the open string theory is admittedly 
one of the most interesting aspects of the proposed open/closed string duality. One expects that the full non-abelian physics
is captured in the bulk by the most general asymptotically flat $p$-brane solution. In this context the virtues of the 
near-horizon limit are well-known and much studied in the context of the AdS/CFT correspondence. 
In this paper we work outside the near-horizon limit and we have chosen to focus on the abelian part around 
a trivial non-abelian vacuum, which provides a comparatively more tractable situation. Yet, we can easily imagine 
more complicated cases where non-abelian effects play a prenounced role.

\vskip 0.1cm
\item{~} From the point of view of the long-wavelength expansions, we can imagine a 0th-order supergravity solution 
that captures holographically a vacuum with non-trivial non-abelian properties. 
It is possible to extract interesting information about non-abelian physics by studying the abelian sector around this 
more general vacuum with a suitable modification of the above-mentioned long-wavelength expansions
in supergravity.

\vskip 0.1cm
\item{~} One example would be to consider an extremal 0th-order solution with a non-vanishing non-abelian condensate. 
The analysis of section \effective\ suggests that the supergravity ansatz is now more complicated with additional 
degrees of freedom and additional interactions. Indeed, as an illustration, deformations of multi-center solutions do 
exhibit these features. It would be very interesting to analyse such examples in more detail, and to attempt to probe 
further our open/closed string conjecture in this direction. Classifying 0th-order solutions by their near-horizon AdS/CFT 
description might prove a useful approach.

\vskip 0.1cm
\item{~} Another example involves $p$-brane solutions at finite temperature. 
Around such vacua the abelian open string effective theory incorporates thermal non-abelian effects and is no
longer the standard DBI action. Typically, it is hard to compute these effects explicitly on the open string side. 
On the gravitational side, however, we are instructed to repeat the blackfold derivative expansion around a 
finite-temperature 0th-order solution. The blackfold equations provide a relatively easy way to compute the 
corresponding modifications of the DBI equations of motion. For instance, finite temperature modifications of 
the DBI action have been considered in this way in \refs{\GrignaniIW\ArmasBK\GrignaniEWA-\GiataganasMLA}. 
In the case of finite temperature configurations at the trivial abelian vacuum the low-energy/near-horizon limit 
of the resulting effective theory is expected to reduce to the fluid dynamical effective description of non-abelian 
dynamics that is familiar from the fluid-gravity correspondence. A related illustration of this statement appeared in 
\EmparanILA.

\bigskip
{\vbox{{\epsfxsize=110mm
\nobreak
\centerline{
\epsfbox{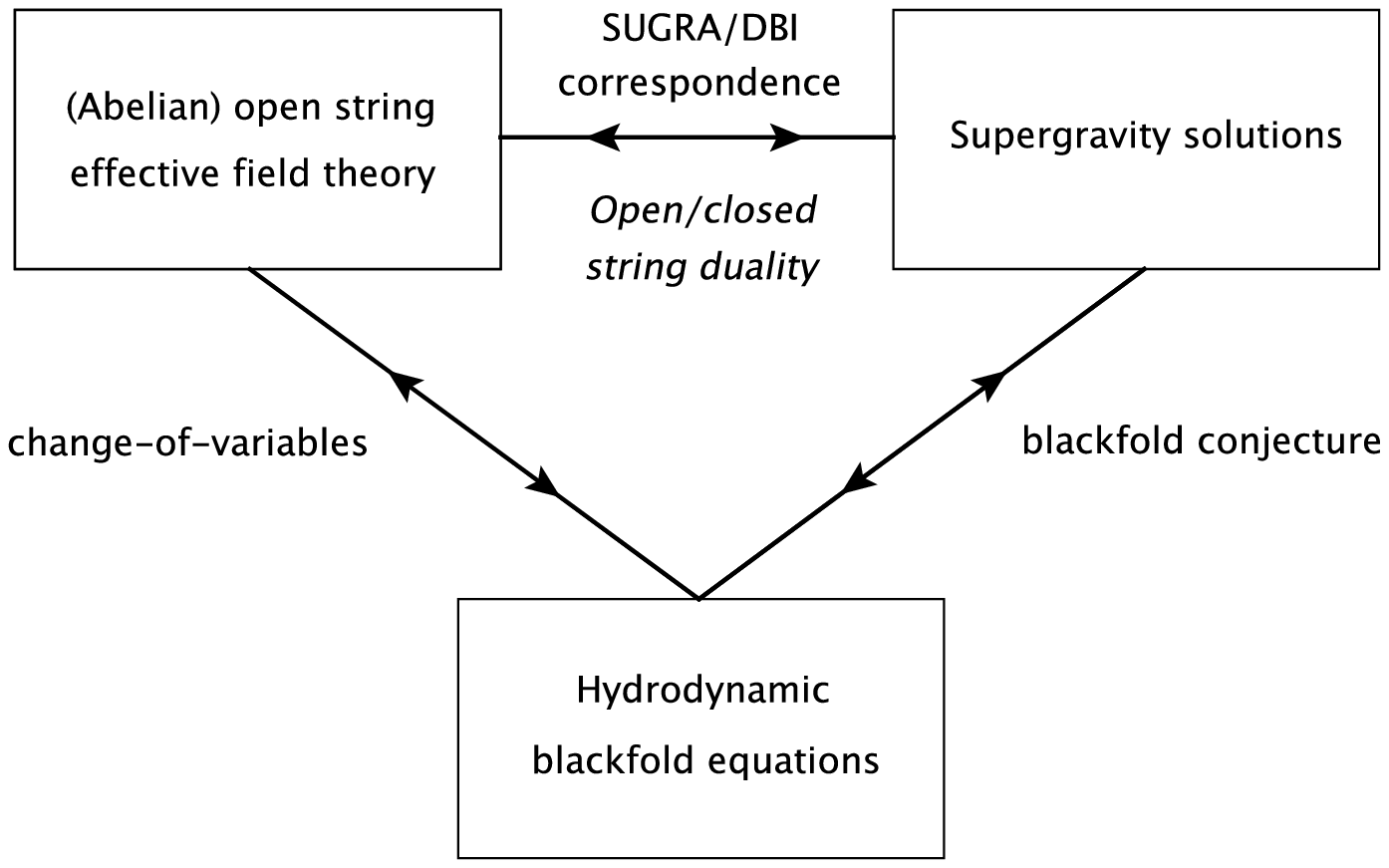}
}
\nobreak
\bigskip
{\raggedright\it \vbox{
{\bf Figure 1.}
{\it A diagrammatic depiction of the main components of our open/closed string conjecture in a long-wavelength derivative expansion.
} }}}}
\bigskip}

\subsec{Open/closed string duality in a long-wavelength regime}
\subseclab\tomography

Closing this section it is useful to summarize some of the most prominent features of the above discussion
highlighting some obvious parallels with the discussion of open string completeness by Ashoke Sen.
We continue to focus on the long-wavelength regime of the proposed open/closed string duality and the 
abelian sector at the origin of the Coulomb branch. 

In the long-wavelength regime we are postulating a picture where the blackfold equations act as a useful 
mediator between gravity and open string theory. This occurs in the way schematically summarized in Fig.\ 1. 
Within the blackfold expansion scheme a part of the supergravity equations reduces to a set of lower-dimensional equations of motion for the collective modes of the supergravity solution, which are naturally formulated as 
hydrodynamic equations. We will show explicitly in the ensuing sections (for extremal configurations) that 
these equations can be reformulated as equations of motion of a recognizable dual open string effective action. 
This direct link between gravity and open string theory is represented by the left arrow, labelled 
{\it change-of-variables}, in Fig.\ 1.

The second ingredient of the construction (right arrow in Fig.\ 1) is the conjecture mentioned above,
which we dub the {\it blackfold conjecture}, stating that solutions of the blackfold equations are in one-to-one 
correspondence with a class of regular $p$-brane solutions. 

Combining these two ingredients we obtain a manifest long-wavelength realization of the proposed open/closed 
string duality: solutions of the abelian action are in one-to-one correspondence with brane supergravity solutions. 
In this form, we set the stage for an explicit algorithmic formulation of the long-anticipated supergravity/DBI 
correspondence, which has been implicit in many previous investigations of brane solutions in supergravity.

In the beginning we argued that it is natural to view this construction as a large-$N$ manifestation of Sen's open 
string completeness. As an obvious similarity with the examples analyzed by Sen \SenNF, we note that 
the abelian open string quantum effective action that we consider in the context of blackfolds is, as in \SenNF, a natural 
description of configurations at energies of the order of the D-brane tension, namely energies of the order $\OO(1/g_s)$. 
The production of closed strings is large and the gravitational solution is deformed, but we postulate that 
the open string description is in itself self-consistent and there is no need to consider a coupled system of 
open and closed strings. In the blackfold derivative expansion this statement is related to the claim that constraint 
equations can be phrased as equations for the collective modes in a fixed supergravity background at all 
orders in the perturbative expansion.

More generally, in complete analogy with Sen's proposal, we do not attempt to set up a holographic relation
in terms of a universal boundary theory that captures all possible gravity (closed string theory) configurations 
in the bulk, but rather we set up holography as {\it a tomographic principle that works in superselection sectors}. 
Different open string theories capture holographically different subsectors in closed string configuration space. 
An open string theory can only capture those closed string configurations that are sourced by the D-brane setup 
on which the open string theory resides. 

This picture has interesting implications for closed string theory and gravity as 
quantum mechanical theories. A viewpoint that pronounces the role of open strings suggests that we should 
separate closed string solutions into different quantum mechanically consistent subsectors, each one with its own
non-perturbative microscopic definition in terms of a dual open string theory. On the other hand, a 
viewpoint centered around a putative single quantum mechanically consistent formulation of closed strings 
and gravity would suggest that we should view closed string theory as an overarching theory-of-theories for diverse 
quantum open string theories (and related quantum field theories) in different subsectors.

\newsec{Relativistic fluids as a link between open and closed strings}
\seclab\link

Our next task is to exhibit the direct relation between the blackfold equations 
(derived from gravity as collective mode equations) and the abelian Dirac-Born-Infeld action, 
making explicit the `change-of-variables' link depicted in Fig.\ 1. We focus on extremal configurations.

We noted already that the blackfold equations are naturally formulated as conservation 
equations of a set of currents which are functionals of the collective modes. With specific 
constitutive relations provided by the thermodynamics of the 0th-order solution, these equations are automatically
formulated within gravity as hydrodynamic equations for a fluid that propagates on an elastic medium \EmparanAT. 
In generic situations of multi-charge $p$-branes this is an anisotropic fluid parametrized by several conserved 
charges, equivalently chemical potentials \EmparanHG. As we will verify soon in explicit examples, the resulting 
hydrodynamic description is non-trivial even at zero temperature if there are non-vanishing chemical potentials.

The passage from hydrodynamics to the DBI theory, described by the change-of-variables arrow in Fig.\ 1, is also 
interesting for another reason. The long-standing problem of Lagrangian reformulations of hydrodynamic systems 
has been revisited recently in several works with promising results, e.g.\ \refs{\DubovskySJ\HaehlPJA-\deBoerIJA}. 
Our analysis provides a different example of such a reformulation. At leading order in the derivative expansion, 
where we encounter ideal relativistic fluids, we discover a reformulation of hydrodynamics in terms of 
a gauge theory with a generalized BF-type interaction. 
At higher orders in the derivative expansion the expected connection to 
open string theory predicts hydrodynamics with specific higher-derivative corrections. In the examples that 
we consider, we find (see section \higherderiv\ below) that these corrections are different in superstring theory compared 
to the bosonic string, and do not have any second order dissipative terms (as one might have anticipated from extremal 
systems).

\newsec{Standard ideal relativistic fluids}
\seclab\fluidreview

Before delving into the details of the hydrodynamic systems that arise in (super)gravity expansions it will be useful
first to set some useful notation and quickly remind the reader of some pertinent facts from the theory of ideal relativistic fluids.  
This topic is very familiar (for a review we refer the reader to \refs{\AnderssonNR,\hydrobook}, and \JackiwNM), 
and does not warrant a special introduction. We slightly generalize the typical setup and consider
$(p+1)$-dimensional ideal fluids on a {\it dynamical} hypersurface propagating in an ambient 
$(d+1)$-dimensional spacetime. We will denote the metric of the ambient spacetime
by $g_{\mu\nu}$ ($\mu,\nu=0,1\ldots,d$), and the induced metric on the fluid hypersurface
by $\gamma_{ab}=g_{\mu\nu} \d_a X^\mu \d_b X^\nu$ ($a,b=0,1,\ldots,p$).\foot{Throughout
the paper we will use small greek letters $\mu,\nu,\ldots$ for the ambient spacetime indices,
and small latin letters $a,b,\ldots$ for indices of the fluid hypersurface. The determinant of the
induced metric $\det(\gamma_{ab})$ will be denoted as $\gamma$.} $X^\mu$ are the 
embedding scalars of the $(p+1)$-dimensional hypersurface inside the $(d+1)$-dimensional
spacetime. To recover the fluid on a fixed background geometry we can simply freeze the dynamics of these
scalars.

It is well known that the equations of motion of an {\it irrotational}\foot{The extention beyond irrotational fluids
is also known and requires replacing $\d_a \theta$ with $\d_a \theta + \alpha\, \d_a \beta$ in 
the action given here, where $\alpha$ and $\beta$ are extra fields. 
We will focus on the irrotational case that is most relevant for our purposes below.} 
relativistic ideal fluid can be formulated as the Euler-Lagrange equations of the action
\eqn\idealaa{
S= \int d^{p+1}x\, \sqrt{-\gamma} \Big[ 
J^a \d_a \theta + f\left( \sqrt{-J^a J_a} \right) 
+b_a \left( J^a -\rho \, u^a \right)
+\lambda \left( u^a u_a +1 \right) \Big]
~.}
The fields $b_a$, $\lambda$ are Lagrange multipliers enforcing the standard relation between 
the fluid current $J^a$, the fluid density $\rho$ and the unit time-like fluid velocity vector $u^a$.
The equations of motion of $\rho$ and $u^a$ set the on-shell
values of the Lagrange multipliers $\lambda=0$, $b_a=0$, and
the function $f$ is an arbitrary function that controls the precise equation of state of the fluid 
(in a manner to be specified momentarily). 

There are three remaining equations of motion. 
The first one comes from the variation of the current $J^a$ 
\eqn\idealab{
\d_a \theta = \frac{J_a}{\sqrt{-J^2}} f'\left(\sqrt{-J^2} \right)
~,}
where $f'$ denotes the derivative of $f$ with respect to its argument and $J^2 = J^a J_a$. Indices are lowered
and raised with the use of the induced metric $\gamma_{ab}$. In differential form language this equation implies 
\eqn\idealaba{
d\left( \frac{f'\left( \sqrt{-J^2} \right)}{\sqrt{-J^2}} \, J \right) =0
~.}
$d$ is the exterior derivative and $J$ the current one-form.

The second equation of motion, that follows from the variation of $\theta$, provides the conservation 
equation of the current
\eqn\idealac{
D_a J^a =0 ~~, ~{\rm equivalently}~~ d*J=0
~.}
$D_a$ is the covariant derivative with respect to $\gamma_{ab}$. 
$*$ denotes the $(p+1)$-dimensional Hodge dual of a differential form with respect to the same metric.

Finally, we can vary the embedding scalars $X^\mu$ that express the induced metric. 
The resulting equations \EmparanAT\ can be massaged into the form
(due to Carter \CarterWV)
\eqn\idealad{
K_{ab}^{~~\hat i}T^{ab}=0
~,}
where $T^{ab}$ is the energy-momentum tensor of the fluid and $K_{ab}^{~~\hat i}$ is the extrinsic curvature tensor.
The latter is expressed in terms of the second derivatives of the embedding scalars (for explicit formulae see for example 
\EmparanAT), and $\hat i$ is a spacetime index along directions perpendicular to the fluid hypersurface.

The energy-momentum tensor of the fluid is (after the use of the ($\lambda, \rho, u^a, b_a$) equations of motion)
\eqn\idealae{
T^{ab} =\frac{2}{\sqrt{-\gamma}} \frac{\delta S}{\delta \gamma_{ab}}
= \left( \varepsilon +P \right)u^a u^b +P \gamma^{ab}
}
with energy density 
\eqn\idealafa{
\varepsilon = f(\rho)
}
and pressure
\eqn\idealafb{
P= \rho f'(\rho) - f(\rho)
~.}
We can see that the arbitrary, but given from the start, function $f$ controls the equation of state of the fluid.
Equations \idealafa, \idealafb\ constitute the standard equation of state of relativistic fluids
that guarantees constant specific entropy \vanHolten. It is straightforward to verify that the equations of motion 
\idealaba\ and \idealac\ imply the conservation of the energy-momentum tensor $T^{ab}$.

Equivalently, it is common to summarize the full set of fluid equations as the conservation 
equations\foot{The elastic equation $K_{ab}^{~~\hat i}T^{ab}=0$ can also be formulated as 
energy-momentum conservation with an index in directions transverse to the fluid hypersurface \CarterWV.}
\eqn\idealag{
D_a T^{ab}=0~, ~~
K_{ab}^{~~\hat i} T^{ab}=0
~.}
Since we are considering irrotational flow we need to supplement these equations with
the irrotational flow condition \idealaba
\eqn\idealai{
d\left( \frac{f'(\rho)}{\rho} J \right) =0
~,}
which is only partially encoded by the energy-momentum conservation condition.

\newsec{Duality in $2+1$ dimensions as a map from fluid dynamics to gauge theory}
\seclab\duality

Actions of the form \idealaa\ have a close relation with classical actions of abelian gauge theories. 
This relation is most straightforward in $2+1$ dimensions. In that case we can dualize the current 
$J^a$ into a 2-form $F_{ab}$
\eqn\dualaa{
F_{ab} =\sqrt{-\gamma}\, \varepsilon_{abc} J^c~~, ~{\rm or ~equivalently}~~ F=*J
~,}
where $\varepsilon_{abc}$ denotes the Levi-Civita antisymmetric symbol $(\varepsilon_{012}=1$).
Then, we notice that the current conservation equation \idealac\ translates into the Bianchi identity
\eqn\dualab{
dF=0
}
and for that reason we can re-interpret $F$ as the field-strength of an abelian 
$(2+1)$-dimensional gauge field $A$, namely $F=dA$.

With these specifications we can reformulate the ideal relativistic fluid \idealaa\ in terms of a modified
abelian Yang-Mills action on a dynamical membrane
\eqn\dualac{
S= \int d^{2+1}x\, \sqrt{-\gamma} \Bigg[
f \left( \sqrt{  \frac{1}{2} F^{ab}F_{ab} }\right) + B^{ab}\left( F_{ab}- \rho\, \varepsilon_{abc}u^c\right) 
+\lambda\left( u^a u_a +1 \right) \Bigg]
~.}
Having assumed \dualaa, \dualab,
the term $J^a \d_a \theta$ is now a total derivative and up to surface terms it does not contribute.
$B^{ab}$ is the Hodge dual of the Lagrange multiplier $b_a$, i.e.\ $b=*B$.

The Euler-Lagrange equations of the action \dualac\ provide an alternative derivation of the ideal fluid 
equations of the previous section (in $2+1$ dimensions). 
In particular, the gauge field equations reproduce the equation \idealaba,
which was closely tied to the irrotational nature of the fluid. 

The reader should appreciate that because of the Lagrange multiplier $B^{ab}$ our system is 
not simply an abelian gauge theory; it is an abelian gauge theory with a specific magnetic 
ansatz for the gauge field
\eqn\dualad{
F_{ab}= \rho \, \varepsilon_{abc} u^c
~.}
Because of the $BF$ coupling we can also view \dualac\ as a generalized BF-type gauge theory.

It is probable that this simple gauge theory reformulation of ideal relativistic hydrodynamics
in $2+1$ dimensions has been noticed before, but I am not aware of an explicit presentation of 
this observation in the literature. Clearly, the duality with a gauge field is specific to $2+1$ dimensions
and would not work in exactly the same manner in other dimensions. Nevertheless, we will soon see that there
other types of fluids that have a close connection with abelian gauge theory in arbitrary spacetime
dimensions.

\bigskip
\noindent
{\bf Special case I : Maxwell theory} 

\medskip
In the special case, where $f(\rho)=\frac{1}{8} \rho^2$, we obtain the ordinary Maxwell theory with a 
magnetic ansatz. In the Maxwell case the energy density and pressure are both positive and equal,
$\varepsilon = P = \frac{1}{8} \rho^2$.

\bigskip
\noindent
{\bf Special case II : Dirac-Born-Infeld theory} 

\medskip
Another interesting case with a different equation of state,
\eqn\dualba{
\varepsilon= - \frac{c^2}{P} 
~,}
where $c$ is an arbitrary constant and the pressure $P$ is negative, can be obtained by setting 
\eqn\dualbb{
f(\rho)= \sqrt{c^2+ \rho^2}
~.}

Because of the identity
\eqn\dualbba{
\frac{1}{c}\det \left( c\, \delta^a_b + F^a_{~b} \right) = c^2 +\frac{1}{2} F^{ab}F_{ab}
}
in three dimensions,
one can easily show that the fluid equations of motion in the case at hand can be written equivalently
as the Euler-Lagrange equations of the Dirac-Born-Infeld action 
\eqn\dualbc{
S= \int d^{2+1}x\, \sqrt{-\gamma} \Bigg[
c \sqrt{\det \left( \delta^a_{~b}+ \frac{1}{c} F^a_{~b}\right)}
+ B^{ab}\left( F_{ab}- \rho\, \varepsilon_{abc}u^c\right) 
+\lambda\left( u^a u_a +1 \right) \Bigg]
~.}

In type IIB string theory this action (with the appropriate overall rescaling) has a familiar connotation: it 
describes the abelian dynamics of {\it extremal} D2-branes with dissolved D0 charge
flowing along the velocity vector $u^a$ inside the D2-brane worldvolume.
In the next section we will re-encounter this fluid from a rather different point of view, that of the supergravity 
(blackfold) analysis of the extremal D0-D2 solution.

\newsec{DBI reconstruction from supergravity}
\seclab\dbisugra

After this short detour on relativistic hydrodynamics we return to the long-wavelength expansions of interest 
in supergravity. To exhibit the relation with the DBI action we will consider in detail two representative 
examples. The first is based on the D0-D2 bound state that we encountered already in the previous section.
The second example is based on the analysis of the F1-D$p$ bound state in flat space.

We will see that different bound states reconstruct the DBI action in different subclasses of configurations.
In this sense the DBI action arises as a master action for whole families of effective hydrodynamic descriptions that
arise from supergravity.

\subsec{D0-D2 deformations}

\noindent
{\it Supergravity analysis}
\medskip

The homogeneous planar D0-D2 bound state in ten-dimensional flat space, that forms the 0-th order solution in 
the long-wavelength expansions of interest, is \refs{\BreckenridgeTT,\CostaZD}
\eqn\ddaa{\eqalign{
ds^2 &= \left( - H^{-\frac{1}{2}} f u_a u_b + D H^{-\frac{1}{2}} \left( \gamma_{ab}+ u_a u_b\right) \right) d\sigma^a d\sigma^b
+H^{\frac{1}{2}}\left( dr^2 + r^2 d\Omega_{6}^2 \right)
,
\cr
e^{2\phi} &= H^{\frac{1}{2}}~,
\cr
B_2 &= \tan\vartheta \, \left( H^{-1}D -1\right) \, * u
~,
\cr
C_{1} & =  \sin\vartheta \coth\alpha \left( H^{-1} - 1 \right) \, u
~,
\cr
C_{3} & = \sec\vartheta \, \coth\alpha \, \left( H^{-1}D -1\right) * {\bf 1}
~,
}}
where 
\eqn\ddab{
f(r) = 1-\left(\frac{r_0}{r}\right)^5~, ~~ 
H(r) = 1+ \left( \frac{r_0}{r} \right)^5 \sinh^2 \alpha~, ~~
D^{-1}(r) = \cos^2\vartheta + \sin^2\vartheta \, H^{-1}
~.}
The solution, which has been boosted with a general $SO(1,2)$ transformation along the $2$-brane worldvolume
coordinates $(\sigma^0, \sigma^1 , \sigma^2)$, is parametrized by the scalars
\eqn\fdac{
\vartheta \in [0,2\pi)~, ~~ \alpha \in \IR~, ~~ r_0 \in \IR_+
~,
}
and the unit velocity vector 
\eqn\ddad{
u^a~, ~~ u_a u^a =-1
~.}

These parameters constitute part of the collective modes of the solution. 
In the metric and $B_2$, $C_1$, $C_3$ potentials we can also see collective coordinates
associated with the breaking of the transverse $SO(7)$ symmetry, which appear through the induced metric 
$\gamma_{ab}$ (to which the Hodge star $*$ refers). These collective modes comprise of 
$7$ transverse scalars $X^{\hat i}$. As usual, it is more convenient to formulate these modes more 
covariantly using ten scalars $X^\mu$ in terms of which the induced worldvolume metric takes the form 
\eqn\ddaea{
\gamma_{ab} = g_{\mu\nu} \d_a X^\mu \d_b X^\nu
~.}
In our problem $g_{\mu\nu}$ is the background Minkowski metric $\eta_{\mu\nu}$. In the planar solution \ddaa\ 
the transverse scalars have a fixed constant value and the induced worldvolume metric is the $3$-dimensional 
flat space metric $\eta_{ab}$. Writing explicitly $\gamma_{ab}$ instead of $\eta_{ab}$ in \ddaa\ is useful, because it prepares us for the type
of supergravity ans\"atze that lead to deformed brane solutions \NiarchosMAA.

In the extremal limit, which will be the main case of interest in this paper, 
\eqn\ddaf{
r_0 \to 0~, ~~ \alpha \to \infty~, ~~{\rm with}~~ r_H^n := r_0^n\, \sinh^2\alpha~~{\rm held~fixed}
~.}
In this limit the temperature vanishes and the solution is 1/2-BPS. 

As explained in previous sections, we want to setup a derivative expansion in supergravity where inhomogeneous 
extremal (but not necessarily supersymmetric) solutions are constructed order-by-order around \ddaa. The supergravity
ansatz is based on the promotion of the above-mentioned collective coordinates to slowly-varying functions
of the worldvolume coordinates $\sigma^a$ \refs{\EmparanCS,\EmparanAT}. Analyzing the constraint equations of 
supergravity within this ansatz at leading order in the derivative expansion one arrives at the extremal blackfold equations
\FluxBlackfolds
\eqn\ddaga{
D_a T^{ab} = 0~, ~~ K_{ab}^{~~\hat i} T^{ab}=0
~,}
\eqn\ddagb{
d* \widetilde J =0 ~, ~~ d*J = 0~, ~~ d*J_3=0
~,} 
with currents
\eqn\ddagc{
T_{ab}= - \CC r_H^5 \left(- \sin^2 \vartheta \, u_a u_b + \cos^2 \vartheta \, \gamma_{ab} \right)~, ~~
\CC := \frac{5\Omega_{(6)}}{16\pi G}
~,}
\eqn\ddagd{
J = \CC \, \sin\vartheta \, r_H^5 \, u~, ~~
\tilde J = \cos\vartheta \, * J
~,}
\eqn\ddage{
J_3 = \CC \, \cos\vartheta \, r_H^5 \, * {\bf 1}
~.}
$\Omega_{(d)} = \frac{2\pi^{\frac{d-1}{2}}}{\Gamma(\frac{d-1}{2})}$ is the volume of the unit round $d$-sphere.
The presence of the 1-, 2-, 3-form currents $J,\tilde J, J_3$ is closely related to the fact that in \ddaa\ the solution
sources the 1-, 2-, 3-form potentials $C_1$, $B_2$, $C_3$.

We notice that the last equation, $d*J_3=0$, in \ddagb\ is expressing trivially the fact that the $2$-brane charge,
which is a quantized quantity, is a worldvolume constant
\eqn\ddagf{
\d_a \left( r_H^5 \, \cos\vartheta \right) =0 ~~ \Leftrightarrow~~ r_H^5 = \frac{c}{\CC \cos\vartheta}
~.}
Here $c$ is an integration constant whose precise value will not play an important role in the ensuing. 
Consequently, $r_H$ is not a true collective mode, and should be substituted in eqs.\ \ddaga, \ddagb\ in terms of 
$\vartheta$. Eq.\ \ddagf\ is consistent with the proposal that open/closed string duality works within superselection
sectors. The superselection sectors in this case are labeled by the value of the integration constant $c$.

To summarize, after the substitution \ddagf\ 
the leading order collective mode (blackfold) equations for D0-D2 solutions deduded from supergravity 
are comprised of the energy-momentum conservation equations
\eqn\ddai{\eqalign{
&D_a T^{ab} = 0~, ~~ K_{ab}^{~~\hat i} T^{ab}=0~,
\cr
&T_{ab}= \left( \varepsilon + P \right) u^a u^b 
+P \gamma_{ab}~,~~
}}
with energy $\varepsilon$, and pressure $P$
\eqn\ddaia{
\varepsilon(\vartheta) = c\, (\cos\vartheta)^{-1} ~, ~~ P(\vartheta) = - c\, \cos\vartheta
~,}
and the current equations
\eqn\ddaj{\eqalign{
&d(\cos\vartheta\, J) =0 ~, ~~ d*J = 0~,
\cr
&J =\rho \, u
~}}
with charge density 
\eqn\ddak{
\rho = c  \, \tan\vartheta
~.}

This is a complete set of dynamical equations for the unknown functions $\vartheta, u^a, X^\mu$. 
The {\it blackfold conjecture} states that solutions of this system are in one-to-one correspondence with 
regular 1st-order corrected inhomogeneous D0-D2 solutions within the blackfold ansatz.\foot{This is 
a natural generalization of the corresponding statement in the fluid-gravity correspondence for large
AdS black holes.} 
In higher-orders of the derivative expansion the conserved currents are corrected with higher-derivative terms,
but the background asymptotic geometry $g_{\mu\nu}=\eta_{\mu\nu}$ is 
not modified. We will discuss higher-order corrections in section \higherderiv.

\bigskip
\noindent
{\it Equivalence with DBI}

\medskip

It is evident that the collective mode equations of the supergravity analysis reduce at leading order 
to the hydrodynamic equations of a standard charged ideal fluid on a dynamical 
surface similar to the one reviewed in section \fluidreview. The constitutive relations are those of an extremal irrotational fluid with 
\eqn\ddba{
\varepsilon = f(\rho) = \sqrt{c^2+ \rho^2}
~.}
The irrotational flow condition \idealai\ coincides with the supergravity equation $d(\cos\vartheta\, J)=0$ in \ddaj.
As we noted in section \fluidreview\ not all equations in \ddai, \ddaj\ are independent. It is enough to consider the equations
\ddai\ combined with the irrotational flow condition. 

In this case the relation to DBI is explained at the end of section \duality\ as special case II. The abelian duality 
of the current $J$ to a gauge field strength \dualaa\ converts the hydrodynamic equations (which are part of the 
supergravity equations here) to the equations of motion of the DBI action with Lagrange multipliers \dualac. 
The basic gauge/gravity map is expressed by equation \dualad\ that translates an open string degree of 
freedom ---the $U(1)$ gauge field with a magnetic ansatz--- to the quantities $\rho$, $u^a$ that express 
components of the gravitational fields.

\subsec{F1-D$p$ deformations}

We now move to a different example in a $(p+1)$-dimensional open string theory $(p>1)$ with an electric gauge field
explicitly turned on.

\bigskip
\noindent
{\it Supergravity analysis}
\medskip

Our starting point, as a 0-th order solution in our long-wavelength expansion,
is the homogeneous, planar F1-D$p$ bound state in ten-dimensional flat space. 
This involves the following supergravity profiles \refs{\CostaZD,\HarmarkWV}
\eqn\fdaa{\eqalign{
ds^2 &= \left( D^{-\frac{1}{2}} H^{-\frac{1}{2}} \left( (1-f) u_a u_b +\hat h_{ab} \right) 
+ D^{\frac{1}{2}}H^{-\frac{1}{2}} \hat \perp_{ab}\right) d\sigma^a d\sigma^b
+D^{-\frac{1}{2}} H^{\frac{1}{2}}\left( dr^2 + r^2 d\Omega_{8-p}^2 \right)
,
\cr
e^{2\phi} &= D^{\frac{p-5}{2}} H^{\frac{3-p}{2}}~,
\cr
B_2 &= \sin\vartheta\, \coth\alpha \, \left( H^{-1}-1\right) \, u\wedge v
~,
\cr
C_{p-1} & = (-)^p \tan\vartheta \left( H^{-1}D-1 \right) \, * (u\wedge v)
~,
\cr
C_{p+1} & = (-)^p \cos\vartheta \, \coth\alpha \, D\left( H^{-1}-1\right) * {\bf 1}
~,
}}
where
\eqn\fdab{
f(r) = 1-\left(\frac{r_0}{r}\right)^n~, ~~ 
H(r) = 1+ \left( \frac{r_0}{r} \right)^n \sinh^2 \alpha~, ~~
D^{-1} = \cos^2\vartheta + \sin^2\vartheta \, H^{-1}~, ~~
n=7-p
~.}
As in the previous section,
the solution has been boosted with a general $SO(1,p)$ transformation along the $p$-brane worldvolume
coordinates $(\sigma^0, \ldots , \sigma^p)$. Accordingly, it is parametrized by the scalars
\eqn\fdac{
\vartheta \in [0,2\pi)~, ~~ \alpha \in \IR~, ~~ r_0 \in \IR_+
~,
}
and the orthogonal vectors 
\eqn\fdad{
u^a~, ~~ v^a~, ~~ (u_a u^a =-1~,~~ v_a v^a =1~, ~~ v_a u^a =0)
~,}
that define the projectors
\eqn\fdae{
\hat h_{ab} = - u_a u_b + v_a v_b~, ~~ 
\hat \perp_{ab} = \gamma_{ab} - \hat h_{ab}
~.}
The induced metric $\gamma_{ab}$, which is flat in the 0-th order solution, incorporates the dependence on
the $9-p$ transverse scalars $X^{\hat i}$. The Hodge star $*$ is defined with respect to this metric.

Once again, we will focus on the extremal limit 
\eqn\fdaf{
r_0 \to 0~, ~~ \alpha \to \infty~, ~~{\rm with}~~ r_H^n := r_0^n\, \sinh^2\alpha~~{\rm held~fixed}
}
where the temperature vanishes and the 0-th order solution is 1/2-BPS. 

Re-analyzing the leading constraint equations in supergravity 
one arrives at the extremal blackfold equations \FluxBlackfolds
\eqn\fdaga{
D_a T^{ab} = 0~, ~~ K_{ab}^{~~\hat i} T^{ab}=0
~,}
\eqn\fdagb{
d*J_{p-1} =0 ~, ~~ d*J_2 = 0~, ~~ d*J_{p+1}=0
~,} 
with currents
\eqn\fdagc{
T_{ab}= - \CC_n r_H^n \left( \sin^2 \vartheta \, \hat h_{ab} + \cos^2 \vartheta \, \gamma_{ab} \right)~, ~~
\CC_n := \frac{n \Omega_{(n+1)}}{16\pi G}~, ~~ n=7-p
~,}
\eqn\fdagd{
J_2 = \CC_n \, \sin\vartheta \, r_H^n \, u\wedge v~, ~~
J_{p-1} = \cos\vartheta * J_2
~,}
\eqn\fdage{
J_{p+1} = \CC_n \, \cos\vartheta \, r_H^n \, * {\bf 1}
~.}
There are three currents expressed as 2-, $(p-1)$- and $(p+1)$-forms corresponding to three spacetime potentials
of the corresponding degree.

The last equation, $d*J_{p+1}=0$, in \fdagb\ is expressing the fact that the $p$-brane charge 
is a worldvolume constant
\eqn\fdagf{
\d_a \left( r_H^n \, \cos\vartheta \right) =0 ~~ \Leftrightarrow~~ r_H^n = \frac{c}{\CC_n \cos\vartheta}
~.}
We solve it in terms of the integration constant $c$ that captures the quantized D$p$-brane charge. 
Then, $r_H$ is substituted in eqs.\ \fdaga, \fdagb\ in terms of $\vartheta$.

After this substitution the leading order F1-D$p$ blackfold equations are
\eqn\fdai{\eqalign{
&D_a T^{ab} = 0~, ~~ K_{ab}^{~~\hat i} T^{ab}=0~,
\cr
&T_{ab}= \left( \varepsilon + P_T \right) u^a u^b - \left( P_T -P_L \right) v^a v^b +P_T \gamma_{ab}~,~~
}}
with energy, transverse and longitudinal pressures
\eqn\fdaia{
\varepsilon(\vartheta) = c\, \frac{1}{\cos\vartheta} ~, ~~ P_T(\vartheta) = - c\, \cos\vartheta~, ~~
P_L(\vartheta) = -c\, \frac{1}{\cos\vartheta}  
~,}
and the current conservation equations
\eqn\fdaj{\eqalign{
&d\left( \cos\vartheta\, J_2 \right) =0 ~, ~~ d*J_2 = 0~,
\cr
&J_2 =c  \, \tan\vartheta ~ u\wedge v
~.}}

The {\it blackfold conjecture} states that solutions of this system are in one-to-one correspondence with 
regular 1st-order corrected inhomogeneous F1-D$p$ solutions in the blackfold ansatz.

\bigskip
\noindent
{\it Equivalence with DBI}
\medskip

In this case we obtain a set of dynamical equations for an augmented set of unknown functions $\vartheta, u^a, v^a, X^\mu$.
They are hydrodynamic equations for an anisotropic fluid propagating on a dynamical $(p+1)$-dimensional hypersurface. 
The general fluid of this type obeys the relations \EmparanHG
\eqn\DFafa{
\varepsilon + P_T = \TT s + \mu q ~, ~~ P_T - P_L = \mu q  
}
where $\TT$ is the local temperature, $s$ the entropy density, $\mu$ the string chemical potential
and $q$ the string charge density. As is evident in \DFafa\
the relation $\varepsilon = - P_L$ is a consequence of extremality $(\TT=0)$.

Generalizing the arguments of section \duality\ in a slightly different direction, which is not merely an abelian Hodge duality,
we will now show that the dynamical system \fdai, \fdaia, \fdaj\ is classically equivalent to the equations of 
motion of the DBI action with an electric field constraint imposed by a 2-form Lagrange multiplier
\eqn\DFaa{\eqalign{
S=\int d^{p+1}x\, \sqrt{-\gamma} &\Big[
\sqrt{\det \left( \delta^a_{~b}+ F^a_{~b}\right)}
+ B^{ab}\Big( F_{ab}- \sin\vartheta\, (u_a v_b - v_a u_b ) \Big) 
\cr
&+\lambda_1 \left( u^a u_a +1 \right) 
+\lambda_2 \left( v^a v_a -1\right) +\lambda_3\, u^a v_a
\Big]
~.}}

The electric field ansatz 
\eqn\DFaba{
F= \sin\vartheta\, u\wedge v
}
is the expected DBI description of the F1-D$p$ system.
The specific form of the field strength \DFaba\ arises if we do an arbitrary pointwise spacetime-dependent 
Lorentz transformation of the constant electric field $F_{01}= \sin\vartheta$ that describes the planar, homogeneous
F1-D$p$ solution. $\sin\vartheta$ is being used to express 
the familiar fact that $F$ (being electric) cannot grow larger than the critical value $F_{01}=1$ where the 
determinant inside the square root vanishes.\foot{We restrict $\vartheta \in [-\frac{\pi}{2},\frac{\pi}{2}]$.}
The orthonormality of the vectors $u,v$, which is part of the definition of \DFaba, is enforced by the 
variation of the Lagrange multipliers $\lambda_1,\lambda_2,\lambda_3$.

The remaining equations of motion of the action \DFaa\ are
\eqn\DFaca{
T^{ab} \, K_{ab}^{~~\hat i}=0
~,}
\eqn\DFacb{
D_a \left( \frac{1}{\cos\vartheta} F^{ab} \right) =0
~.}
In addition, we have the Bianchi identity
\eqn\DFacc{
dF=0
~.}
With the ansatz \DFaba\ these equations are obviously the same as the hydrodynamic equations \fdai, \fdaia, \fdaj.

To verify this, first we notice that 
the energy-momentum tensor of this system (after the use of the $\lambda_{1,2,3}$, $u_a, v_a$, $B^{ab}$,
$\vartheta$ equations) is 
\eqn\DFab{
T^{ab} = \frac{\sin^2\vartheta}{\cos\vartheta} (u^a u^b - v^a v^b)
-\cos\vartheta \, \gamma^{ab} 
~}
the same as that encountered in the blackfold equations.

Second, since 
\eqn\DFag{
J_2 = \frac{c}{\cos\vartheta} F
} 
we observe that the gauge field equation \DFacb\ is identical to the string current conservation equation $d*J_2=0$. 
The Bianchi equation \DFacc, which is identical to the first equation in \fdaj, $d(\cos\vartheta\, J_2)=0$,  
can be viewed as a property closely related to the irrotational 
condition \idealaba\ of standard relativistic fluids in section \fluidreview.

\subsec{Extensions}

More general configurations of $p$-brane bound states with dissolved lower-dimensional charges in 
flat space admit a similar analysis. The constraint equations of the perturbative supergravity 
analysis always reduce to a hydrodynamic system, which admits a direct reformulation at extremality as a 
DBI action along the lines described above. 

Let us summarize several interesting extensions of the exercises of the previous two subsections. 
An important extension concerns the higher-derivative corrections. There are such corrections both on the blackfold
supergravity analysis and on the open string side as corrections to the abelian DBI effective action. We will discuss the 
latter in the next section.

Another interesting direction, that was highlighted already in section \OCduality, concerns the incorporation of non-abelian 
effects. The imprint of such effects in the abelian sector and its effective description can be captured in a conceptually 
straightforward manner in the blackfold supergravity approach by re-doing the perturbative analysis around other 
0th-order $p$-brane solutions. 

For instance, we can consider flat $p$-brane bound states at finite temperature. In the above-mentioned D0-D2, 
F1-D$p$ examples this introduces an additional degree of freedom (both $r_0$ and $\alpha$ participate without 
the scaling \fdaf). It is interesting to ask if there is a finite-temperature deformation of the DBI action that reproduces 
the on-shell finite-temperature hydrodynamic equations. This question was considered for stationary 
configurations in Refs.\ \refs{\GrignaniIW\ArmasBK-\GrignaniEWA}. However, the question of a general 
deformation of the DBI action independent of stationarity remains open. Recent advances, e.g.\ in Ref.\ \HaehlPJA, 
could prove a useful avenue for this problem. We should point out that thermal small temperature corrections to 
the DBI action have been computed at weak coupling in Ref.\ \GrignaniEWA, yet the exact form of finite 
temperature corrections in open string theory is hard to obtain.\foot{For a discussion of thermal corrections 
to gauge theory from a D-brane probe analysis in the context of the AdS/CFT correspondence see \KiritsisTX.}

Deformations of the abelian DBI action can also be obtained from supergravity at extremality by considering other exact
$p$-brane solutions at 0th-order. For instance, deformations of the 0th-order solution will appear necessarily under 
external forcing, i.e.\ when a $p$-brane solution is embedded in a non-flat asymptotic background with fluxes. 
As a concrete example consider brane solutions in AdS.\foot{AdS black holes in the blackfold approximation have 
been considered in \refs{\CaldarelliPZ,\ArmasHZ}.}  
In this case the hydrodynamic blackfold equations are modified in two different ways compared to the equations in flat space.  
First, there is a background-induced deformation of the conserved currents. Second, there are extra force terms in the 
equations, which are analogs of the WZ couplings in the DBI action. A general formulation of such couplings in the blackfold 
expansion will appear in \FluxBlackfolds. Further studies of such effects should contribute significantly to the understanding
of the open/closed string duality proposed in section \OCduality.

\newsec{Higher-order hydrodynamics from higher-derivative corrections to DBI}
\seclab\higherderiv

The DBI action is the leading term in a long-wavelength derivative expansion of open string field theory.
Open string theory dictates very specific higher-derivative corrections to the DBI action. 
Such corrections were determined in flat space in \refs{\TseytlinWW\AndreevCB\AbouelsaoodGD-\AndreevRE}
using the S-matrix or $\sigma$-model approach (for a review see \TseytlinDJ).
Since we make a connection with hydrodynamics 
it is interesting to ask how such corrections translate into the hydrodynamic language. 

In what follows we will assume the validity of the non-renormalization relation \openai.
Combined with open/closed string duality this relation allows us to translate information from a weak coupling
open string analysis into a set of predictions for appropriate supergravity solutions in a long-wavelength derivative
expansion.

Before we go into the details of the connection between open string and fluid dynamical higher derivative connections, 
it is useful to recall that 
the subject of higher-derivative (dissipative) corrections in relativistic hydrodynamics has a long history.
The mere addition of 1st order derivative corrections to the energy-momentum tensor and the current
\refs{\Eckart,\LF} is well known to be inadequate and leads to unacceptable problems with 
causality and stability. These problems are amended in the Israel-Stewart approach \refs{\StewartI\StewartII-\StewartIII},
where higher-order corrections are added, or in other formalisms like Carter's canonical 
formalism (for a review see e.g.\ \refs{\AnderssonNR,\hydrobook}). 

The embedding of the DBI action in open string theory and its map to gravity 
suggests a class of hydrodynamic systems with higher-derivative corrections derived from string theory. 
The consistency of the latter implies that the usual issues with causality observed in generic (low order) hydrodynamic
constructions should be absent here. Keep in mind that compared to the generic case discussed in the 
hydrodynamics literature, in this paper we have focused mainly on zero-temperature, finite-density fluids. 
This necessarily entails some obvious differences compared to the standard discussion 
of finite-temperature relativistic fluids that will become apparent soon. 

Following \AndreevCB\ one finds from an open string theory computation that the general form
of the leading higher-derivative corrections to the DBI action in superstring theory in flat space is\foot{We
drop factors of $\pi$ and $\alpha'$ that can be easily reinstatated.}
\eqn\viscaa{
S_{super}= \int d^{p+1} x \, \sqrt{-\det(\eta_{ab}+F_{ab})} 
\Big[ 1+ \FF^{k\ell m n abcd}(F) \, 
\d_k \d_\ell F_{mn} \d_a \d_b F_{cd} 
+ \OO(\d^6)
\Big]
~.}
For simplicity, in \viscaa\ we have frozen the background geometry and the transverse scalar dynamics. 
The transverse scalar dynamics can be derived from this action in ten dimensions by 
T-duality.\foot{It would be interesting to work out these corrections explicitly and compare with the
general theory of relativistic elasticity discussed in \refs{\ArmasJG\ArmasHSA\ArmasGOA\ArmasRVA-\ArmasNEA}.} 
We notice that the leading higher-derivative correction comes at the order of four derivatives. The function 
$\FF(F)\sim F^2 + F^4 +\ldots$ has an in principle computable expansion in powers of the field
strength $F$. For example, a 4-vector superstring amplitude calculation on the disc gives up to order 
$\OO(\d^4 F^6)$ \AndreevCB
\eqn\viscab{\eqalign{
&\FF^{k\ell m n abcd} \d_k \d_\ell F_{mn} \d_a \d_b F_{cd}
= 
\cr
&-\frac{1}{96} \Big( \d_a \d_b F_{mn} \d^a \d^b F^{n\ell} F_{\ell r} F^{rm}
+\frac{1}{2} \d_a \d_b F_{mn} F^{n\ell} \d^a \d^b F_{\ell r} F^{rm} 
\cr
&-\frac{1}{4} \d_a \d_b F_{mn} F^{mn} \d^a \d^b F^{\ell r} F_{\ell r}
-\frac{1}{8} \d_a \d_b F_{mn} \d^a \d^b F^{mn} F_{\ell r} F^{\ell r}
\Big) + \OO(\d^4 F^6)
~.}}

Interestingly, the corrections are different in the bosonic string.\foot{I would like to thank
E.\ Kiritsis for emphasizing this point.} Quoting \refs{\TseytlinWW,\AndreevCB}
\eqn\viscac{
S_{bosonic}=\int d^{p+1} x \, \sqrt{-\det( \eta_{ab} + F_{ab} ) }
\Big[ 1+ \FF^{kmnacd}(F) \d_k F_{mn} \d_a F_{cd} 
+\OO(\d^4)
\Big]
~.}
From a 4-vector amplitude on the disc one finds
\eqn\viscad{\eqalign{
&\FF^{kmnacd}(F) \d_k F_{mn} \d_a F_{cd} 
= -\frac{1}{48 \pi} \Big[
F_{k\ell}F^{k\ell} \d_a F_{mn} \d^a F^{mn} 
\cr
& +8 F_{k\ell} F^{\ell m} \d_a F_{mn} \d^a F^{nk}
-4 F_{\ell a} F^{\ell b} \d^a F^{mn} \d_b F_{mn}
+\OO(\d^2 F^6)
\Big]
~.}}
In this case the derivatives start at a lower order, $\OO(\d^2)$.

The strategy developed in the previous sections suggests a natural connection of these actions 
(supplemented with specific ans\"atze for the gauge field strength) with higher-derivative hydrodynamic systems. 
For example, a D0-D2-type higher-derivative fluid in three dimensions arises from the action 
\eqn\viscae{
S= S_{\rm OS} -\int d^{2+1}x\, \sqrt{-\gamma} \Big[ B^{ab}\left( F_{ab}-\rho\, \varepsilon_{abc}u^c)
+\lambda (u^a u_a +1)\right) \Big]
~,
} 
where $S_{\rm OS}$ is the open string-derived DBI action with higher derivative corrections
($S_{super}$ or $S_{bosonic}$ above). As we noted previously the Lagrange multiplier $B^{ab}$
enforces the ansatz $F_{ab}= \rho \, \varepsilon_{abc}u^c$ and then the Bianchi identity $dF=0$
guarantees the current conservation $d*J=0$, where $J^a =\rho u^a$.
This particular identification of the current (unchanged by the presense of the derivative corrections) 
means that we have chosen to work in the Eckart frame.

A significant part of the gauge field equations can be re-expressed as the energy-momentum conservation conditions
$D_a T^{ab}=0$. As an illustration let us consider the higher-derivative corrections to the energy-momentum tensor
that arise in the case of the bosonic string. After the implementation of the $B^{ab},\lambda, \rho, u^a$ equations the 
energy-momentum tensor of the resulting fluid takes the following form up to $\OO(\d^2)$
\eqn\viscaf{
T^{ab} = T^{ab}_{ideal}+ T^{ab}_{higher-derivative}
}
with 
\eqn\viscag{
T^{ab}_{ideal}
= \frac{1}{\sqrt{1+\rho^2}} \Big( \rho^2\, u^a u^b - \gamma^{ab} \Big)  
} 
and
\eqn\viscai{
T^{ab}_{higher-derivative} = 
\Big( T^{ab}_{ideal}\, \widetilde F^{k\ell}_{mn} 
+\sqrt{1+\rho^2} \left[ \widetilde F^{k\ell}_{mn} \right]^{ab} \Big)
\D_k \left( \rho\, u^m\right) \D_\ell \left( \rho\, u^n \right)
~.}
The tensor structures 
\eqn\viscaj{\eqalign{
&\widetilde F_{mn}^{k\ell}
= \varepsilon_{dem} \varepsilon_{fgn} \FF^{k de \ell fg}~, 
\cr
&\left[ \widetilde F^{k\ell}_{mn} \right]^{ab} \d_k \left( \rho u^m \right) \d_\ell \left( \rho u^n\right) 
= \varepsilon_{dem} \varepsilon_{fgn} 
\frac{\d}{\d \gamma_{ab}}\Big [ \FF^{kdefg\ell} D_k (\rho u^m) D_\ell (\rho u^n) \Big]
}}
are functions of $\rho$, and $u$ without derivatives. At the order of equation \viscad\ we obtain 
the more specific expressions
\eqn\viscak{\eqalign{
&\widetilde F^{k\ell}_{mn} \D_k \left( \rho\, u^m\right) \D_\ell \left( \rho\, u^n \right)
\cr
&=  \frac{\rho^2}{12\pi}
\Big( -3 \d_a \rho \d^a \rho +\rho^2 \, \D_a u_b \D^a u^b
+ u_a \d^a \rho\, u_b \d^b \rho 
-2 \rho^2 u_a \D^a u_c\, u_b \D^b u^c
\Big) 
~,}}
\eqn\viscal{
\left[ \widetilde F^{k\ell}_{mn} \right]^{ab} \D_k \left( \rho\, u^m\right) \D_\ell \left( \rho\, u^n \right)
= \frac{\rho^2}{6\pi} \Big(
\AA^{ab}\left((\d \rho)^2 \right) + \rho\, \BB^{ab} \left( \d \rho \D u \right)
+\rho^2 \,\CC^{ab} ( (\D u)^2)  \Big)
}
where
\eqn\viscala{\eqalign{
\AA^{ab} =& ~
\d^a \rho\, \d^b \rho 
-2 \hat \perp^{ab} u^c \d_c\rho \, u^d \d_ d \rho
- 4 u^{(a} \d^{b)}\rho \, u^c \d_c \rho
+2 \varepsilon^{act} \varepsilon^{bds} u_t u_s \d_c \rho \d_d \rho
,
\cr
\BB^{ab} =&~ 4 \d_c \rho \, \D^c u^{(a} u^{b)} - 4 u^c \d_c\rho\, u^d \D_d u^{(a} u^{b)} 
~,
\cr
\CC^{ab} =& ~
\D_c u^a \D^c u^b 
+ \D^a u^c \D^b u_c 
-3 u^a u^b \D_c u_d \D^c u^d 
-2 u^c \D_c u^a \, u^d \D_d u^b
\cr
&+4 u^{(a} \D^{b)} u^d \, u^c \D_c u_d 
+2 \gamma^{ab} u^c \D_c u_e \, u^d \D_d u^e
+4 \varepsilon^{aks}\varepsilon^{bmt} u_s u_t \D_c u_k \D^c u^m
\cr
&-2 \varepsilon^{act} \varepsilon^{bds} u_t u_s \D_c u_e \D_d u^e
~.}}
To derive these expressions we promoted the worldvolume metric $\eta_{ab}$ to $\gamma_{ab}$
covariantizing all couplings in \viscac, \viscad, then took a derivative with respect to $\gamma_{ab}$ 
and finally set $\gamma_{ab}=\eta_{ab}$. Potential higher derivative couplings of the worldvolume
metric do not affect this computation.

Several comments are in order at this point:

\item{$(i)$} We observe that the usual dissipative corrections at order $\OO(\d)$ associated to 
the shear and bulk viscocity are absent. This is due to the extremal nature of 
the configurations that we are considering. At non-zero temperature such corrections
are expected to appear. In fact we would expect that the ratio of the shear viscocity over the
entropy density, $\frac{\eta}{s}$, is non-vanishing at non-zero temperature. The connection 
with gravity in previous sections suggests that this ratio is the constant $1/4\pi$. Then, for the system
at hand we expect that as we take the zero-temperature limit $\eta/s$ remains non-vanishing 
while both $\eta$ and $s$ go simultaneously to zero.

\item{$(ii)$} The leading corrections occur at $\OO(\d^2)$. Qualitatively these are corrections of 
the same general form as the corrections in the Israel-Stewart formalism \refs{\StewartI\StewartII-\StewartIII}.

\item{$(iii)$} It is interesting to ask how field redefinitions affect the above formulae. For instance,
in a different frame, e.g.\ the Landau frame, where the current $J$ is $\rho u^a + \OO(\d)$ corrections,
the action \viscac\ will also receive derivative corrections from the expansion of the DBI square root.
Notice that the leading corrections remain $\OO(\d^2)$.

\bigskip

Repeating the same exercise with the superstring action \viscaa\ we find a different set of higher-derivative corrections. 
For example, in the case of the D0-D2 configurations the above analysis would yield a fluid whose
energy-momentum tensor receives its leading higher-derivative contributions at $\OO(\d^4)$.
It would be interesting to know if this feature is related to the improved convergence properties of DBI solutions
in superstring theory that have been observed throughout the literature over the years, e.g.\ 
the BIon solutions \CallanKZ.

A similar analysis of open string-derived derivative corrections can be performed for 
the F1-D$p$ configurations (that require the ansatz
$F=\sin\vartheta \, u\wedge v$), or other more general configurations of the gauge field
that lead to anisotropic fluids. We will not spell out the details here.

\bigskip

\centerline{\bf Acknowledgements}

\bigskip
\noindent
I would like to thank E. Floratos, C.\  Rosen, A.\ Sen, T.\ Tomaras and A.\ Tseytlin for useful discussions. 
I am especially grateful to E.\ Kiritsis and N.\ Obers for many useful and illuminating conversations 
on related topics over the years and their comments on a preliminary draft of the paper. Finally, I am indebted to 
J.\ Armas, J.\ Gath, N.\ Obers, and A.\ Pedersen for the collaboration in a related project \FluxBlackfolds, 
from which certain aspects of the present paper benefited greatly. 
This work was supported in part by European Union's Seventh Framework Programme under grant 
agreements (FP7-REGPOT-2012-2013-1) no 316165, PIF-GA-2011-300984, the EU program ``Thales'' 
MIS 375734  and was also co-financed by the European Union (European Social Fund, ESF) and Greek national 
funds through the Operational Program ``Education and Lifelong Learning'' of the National Strategic Reference 
Framework (NSRF) under ``Funding of proposals that have received a positive evaluation in the 3rd and 4th Call 
of ERC Grant Schemes''.

\listrefs
\end